\documentclass[superscriptaddress,preprintnumbers,nofootinbib,twocolumn]{revtex4-2}

\usepackage{hyperref}
\usepackage{tikz-feynman}
\usepackage{subfig}
\usepackage{bm}
\usepackage{esvect}
\usepackage{verbatim}
\usepackage{multirow}

\usepackage[left]{lineno}
\usepackage{blindtext}
\pdfoutput=1
\usepackage[T1]{fontenc}
\usepackage{graphicx}
\usepackage{epsfig}
\usepackage{amsmath}
\usepackage{amsfonts}
\usepackage{amssymb}
\usepackage{braket}
\usepackage{cancel}
\usepackage{pstricks}
\usepackage{color}
\usepackage{feynmf}
\usepackage[normalem]{ulem}
\usepackage{epstopdf}
\usepackage{soul}
\usepackage{cancel}

\makeatletter
\renewcommand\@makecaption[2]{%
  \par
  \vskip\abovecaptionskip
  \begingroup
   \small\rmfamily
    \begingroup
     \samepage
     \flushing
     \let\footnote\@footnotemark@gobble
     \@make@capt@title{#1}{#2}\par
    \endgroup
  \endgroup
  \vskip\belowcaptionskip
}
\makeatother

\definecolor{ao(english)}{rgb}{0.0, 0.5, 0.0}

\usepackage{braket}
\usepackage{cancel}
\usepackage{setspace}
\usepackage{pstricks}
\usepackage{xcolor}
\usepackage{float}
\usepackage{mathtools}
\usepackage{multirow}
\usepackage{booktabs}
\usepackage{enumitem}
\providecommand{\tabularnewline}{\\}

\def\gsim{\lower0.5ex\hbox{$\:\buildrel >\over\sim\:$}}
\def\lsim{\lower0.5ex\hbox{$\:\buildrel <\over\sim\:$}}

\begin{document}

\title{Probing the muon $(g-2)$ anomaly at the LHC \\
in final states with two muons and two taus}
\author{Yoav Afik}
\email{yoavafik@gmail.com}
\affiliation{Experimental Physics Department, CERN, 1211 Geneva, Switzerland}

\author{P. S. Bhupal Dev}
\email{bdev@wustl.edu}
\affiliation{Department of Physics and McDonnell Center for the Space Sciences, Washington University, St.~Louis, MO 63130, USA}

\author{Amarjit Soni}
\email{adlersoni@gmail.com}
\affiliation{Physics Department, Brookhaven National Laboratory, Upton, NY 11973, USA}

\author{Fang Xu}
\email{xufang@wustl.edu}
\affiliation{Department of Physics, Washington University, St.~Louis, MO 63130, USA}

\date{\today}   

\begin{abstract}
 The longstanding muon $(g-2)$ anomaly, as well as some hints of lepton flavor universality violation in $B$-meson decays, could be signaling new physics beyond the Standard Model (SM). A minimal $R$-parity-violating supersymmetric framework with light third-generation sfermions (dubbed as `RPV3') provides a compelling solution to these flavor anomalies, while simultaneously addressing other pressing issues of the SM. We propose a new RPV3 scenario for the solution of the muon $(g-2)$ anomaly, which leads to an interesting LHC signal of $\mu^+\mu^-\tau^+\tau^-$ final state. We analyze the Run-2 LHC multilepton data to derive stringent constraints on the sneutrino mass and the relevant RPV coupling in this scenario. We then propose dedicated selection strategies to improve the bound even with the existing dataset. We also show that the high-luminosity LHC will completely cover the remaining muon $(g-2)$-preferred parameter space, thus providing a robust, independent test of the muon $(g-2)$ anomaly.

\end{abstract}

\maketitle
\flushbottom


\section{Introduction \label{sec:intro}}
The magnetic moment of the muon ($g_\mu$) is one of the most precisely measured quantities in particle physics and an important ingredient to precision tests of the Standard Model (SM)~\cite{Workman:2022ynf}. Intriguingly, the anomalous magnetic moment of the muon, $a_\mu\equiv (g_\mu-2)/2$, arising from loop corrections to the fermionic electromagnetic vertex, was found to have a $3.7\sigma$ discrepancy between the experimental value from the E821 experiment at Brookhaven and the SM prediction~\cite{Muong-2:2006rrc}. The situation became even more interesting recently, as the first result from the Fermilab Muon $(g-2)$ experiment~\cite{Muong-2:2021ojo}, utilizing a more intense muon beam and improved detectors was shown to be consistent with the old Brookhaven measurement to six significant figures. When combined and compared with the world-average of the SM prediction using the ``R-ratio method''~\cite{Aoyama:2020ynm}, the discrepancy increases to $4.2\sigma$: 
\begin{align}
    \Delta a_\mu \equiv a_\mu^{\rm exp} - a_\mu^{\rm SM} = (251 \pm 59) \times 10^{-11} \, .
    \label{eq:deltamu}
\end{align}
It should be noted here that simultaneously with the announcement of the Fermilab result in 2021, a new lattice simulation result from the BMW collaboration was also published~\cite{Borsanyi:2020mff}. The BMW result for the leading hadronic contribution to $a_\mu$ reduces the discrepancy in $\Delta a_\mu$ to only $1.5\sigma$. At that time most other lattice collaborations did not have their results available. This situation has changed now. In the past few months several lattice collaborations have made their results available~\cite{Ce:2022kxy,Alexandrou:2022amy,Colangelo:2022vok,talkLehner2022,talkGottlieb2022,talkColangelo2022} in the ``intermediate distance regime'',   {\it i.e.} from 0.4 to 1.0 fermi. In that intermediate regime, almost all lattice collaborations now seem to agree with BMW. The interpretation of these new lattice results seems to be that the tension with experiment is only of order 3.1$\sigma$, {\it i.e.} somewhat less than the R-ratio method indicated. However, the new lattice results are in some tension with the low energy $e^+e^-\to {\rm hadrons}$ cross-section data~\cite{Crivellin:2020zul, Keshavarzi:2020bfy, Colangelo:2020lcg,  talkColangelo2022}, so further clarification 
is needed. 
In the coming years, more refined lattice results should be forthcoming and are eagerly awaited.
Until all these issues get resolved we choose to use the discrepancy quoted in Ref.~\cite{Muong-2:2021ojo} and shown in Eq.~\eqref{eq:deltamu}.

Taking the muon $(g-2)$ anomaly at face value, one could ask what kind of beyond the SM (BSM) physics might be responsible. The answer is many~\cite{Lindner:2016bgg, Athron:2021iuf,Afik:2021xmi}. The leading one-loop contribution from any new physics (NP) source is parametrically of the order of 
\begin{align}
    a_\mu^{\rm NP}\sim \frac{g_{\rm NP}^2}{16\pi^2}\frac{m_\mu^2}{m_{\rm NP}^2} \, ,
    \label{eq:NP}
\end{align}
which should coincidentally be at the same level as the SM electroweak contribution~\cite{Jackiw:1972jz} 
\begin{align}
    a_\mu^{\rm EW}[{\rm 1-loop}]= \frac{g^2}{16\pi^2}\frac{m_\mu^2}{m_W^2}f\simeq 194.8\times 10^{-11}\, , 
\end{align}
(where $f=\left[5+(1-4\sin^2\theta_W)^2\right]/12\simeq 0.4$) in order to explain the discrepancy in Eq.~\eqref{eq:deltamu}. Hence, there are essentially two types of solutions, depending on whether the new physics contains (i) small couplings and small masses compared to the electroweak scale, as in {\it e.g.} axion, dark photon, and light $Z'$ models; or (ii) ${\cal O}(1)$ interactions and ${\cal O}(100~{\rm GeV})$ masses,\footnote{In some new physics models, the SM-like scaling $ a_\mu^{\rm NP}\propto m_\mu^2$ in Eq.~\eqref{eq:NP} can be avoided by chiral enhancement inside the loop, thus allowing for viable solutions with higher masses up to tens of TeV~\cite{Czarnecki:2001pv,Crivellin:2022wzw, Stockinger:2022ata}.} as in {\it e.g.} two-Higgs doublet, supersymmetry, and leptoquark models~\cite{Lindner:2016bgg, Athron:2021iuf}. There is no restriction on the new particle(s) in the loop contributing to $g-2$, except that in most cases we need to invoke flavor non-universal couplings to avoid other experimental constraints. In this context, the models with a new coupling to the $\mu-\tau$ sector are particularly appealing, because of the relatively weaker constraints involving the tau lepton. We will assume this to be the case for the solution to the muon $(g-2)$ anomaly, and explore how this scenario can be {\it directly} tested at the LHC using final states with two muons and two taus.\footnote{For other interesting ideas on testing the muon $(g-2)$ at colliders, see {\it e.g.} Refs.~\cite{Freitas:2014pua,Sabatta:2019nfg,Capdevilla:2021rwo,Arkani-Hamed:2021xlp}. The same final state was also considered in Ref.~\cite{Babu:2020ivd} in the context of an $SU(2)_H$ model for large neutrino magnetic moments.} 

A particularly attractive BSM scenario is $R$-parity violating supersymmetry (RPV-SUSY)~\cite{Barbier:2004ez}, which has the virtue to address many shortcomings of the SM, such as nonzero neutrino masses, radiative stability of the Higgs boson, radiative electroweak symmetry breaking, stability of the electroweak vacuum, gauge coupling unification,  (gravitino) dark matter and baryogenesis. Here we focus on a minimal, well-motivated RPV-SUSY framework with the third-generation superpartners lighter than the first two, hence dubbed as `RPV3'~\cite{Altmannshofer:2017poe}, which preserves all the attractive features of SUSY mentioned above. On top of that, it was recently shown~\cite{Altmannshofer:2020axr, Dev:2021ipu} that RPV3 can simultaneously explain the muon $(g-2)$ anomaly, along with other persistent hints of lepton flavor universality violation in semileptonic $B$-meson decays, most significantly the $R_{D{^{(*)}}}$ and $R_{K{^{(*)}}}$ anomalies.\footnote{For reviews of the $B$-anomalies and BSM interpretations, see e.g. Refs.~\cite{Fischer:2021sqw,Crivellin:2021sff}. For RPV-SUSY interpretations of the flavor anomalies, see also Refs.~\cite{Deshpande:2012rr,Biswas:2014gga,Zhu:2016xdg,Deshpande:2016yrv,Das:2017kfo,Earl:2018snx, Trifinopoulos:2018rna,Hu:2018lmk,Trifinopoulos:2019lyo,Wang:2019trs,Hu:2019ahp,Zheng:2021wnu,Bardhan:2021adp,Zheng:2022ssr}.}  
Another important feature of the RPV3 solution proposed in Refs.~\cite{Altmannshofer:2020axr, Dev:2021ipu} is that the muon $(g-2)$ anomaly is primarily governed by the $LLE$-type interactions [cf.~Eq.~\eqref{Eq.RPVLLE}], while the $R_{D{^{(*)}}}$ and $R_{K{^{(*)}}}$ anomalies are governed by the $LQD$-type interactions [cf.~Eq.~\eqref{Eq.RPVLQD}]. This mutual orthogonality allows us to explore here the LHC prospects of probing the muon $(g-2)$-preferred parameter space, irrespective of the fate of the $B$-anomalies. 

For the benchmark scenario considered in Ref.~\cite{Dev:2021ipu} with only $\lambda_{232}=-\lambda_{322}\neq 0$ (and all other $\lambda_{ijk}=0$), there is a spectacular four-muon signal at the LHC~\cite{Chakraborty:2015bsk}, coming from the tau sneutrino pair-production, followed by each sneutrino decaying into two muons via the $\lambda_{232}$ coupling. Recasting a recent ATLAS multilepton analysis~\cite{ATLAS:2021eyc}, we obtained a lower bound of $m_{\widetilde{\nu}_\tau}\gtrsim 670$ GeV, which ruled out most of the muon $(g-2)$-preferred region and pushed the $\lambda_{232}$ coupling toward the perturbative limit of $\sqrt{4\pi}$.     

Given the fact the collider signals involving tau final states are in general less constrained than those involving electrons or muons, in this paper we explore a new RPV3 benchmark with $\lambda_{233}=-\lambda_{323}\neq 0$, which leads to a final state with two muons and two taus at the LHC [cf.~Fig.~\ref{fig:signal}]. To the best of our knowledge, there are no existing constraints on sneutrinos that can be directly applied to this scenario (without any additional assumptions), except the model-independent LEP limit of $m_{\widetilde{\nu}_\tau}>41$ GeV from $Z$ invisible decay width measurements~\cite{ALEPH:1991qhf}. Our goal in this paper is to remedy this situation and derive the {\it first direct LHC limit} on sneutrinos for the $\lambda_{233}\neq 0$ case. To this end, we repurpose a recent ATLAS analysis~\cite{ATLAS:2021yyr} to study the $\mu^+\mu^-\tau^+\tau^-$ signal and background at $\sqrt s=13$ TeV LHC with an integrated luminosity of 139 fb$^{-1}$. As a result, we are able to put a {\it new} robust lower limit on $m_{\widetilde{\nu}_\tau}$ extending to about 400 GeV. When contrasted with the muon $(g-2)$-preferred region, we get a conclusion similar to Ref.~\cite{Dev:2021ipu}, {\it i.e.} only large values of $\lambda_{233}$ close to the perturbative limit are compatible with the muon $(g-2)$-anomaly in this scenario. We also give the future projections at the high-luminosity phase of the LHC (HL-LHC), which will be able to completely probe the remaining muon $(g-2)$-preferred parameter space, thus providing an independent probe of the muon $(g-2)$-anomaly.

The rest of the paper is organized as follows: in Section~\ref{sec:RPV3}, we briefly review the salient features of the RPV3 model framework and how it explains the muon $(g-2)$ anomaly. Section~\ref{sec:collider} presents the details of the signal and background analysis for the  $\mu^+\mu^-\tau^+\tau^-$ channel. Our results are summarized in Section~\ref{sec:results}. Section~\ref{sec:conclusions} gives the conclusions. Some additional kinematic distributions are shown in Appendix~\ref{sec:kinematic}.  

\section{Muon $(g-2)$ in the RPV3 Framework}
\label{sec:RPV3}
As suggested earlier~\cite{Altmannshofer:2017poe, Altmannshofer:2020axr, Dev:2021ipu}, the RPV3 framework provides an appealing solution to the flavor anomalies. The relevant pieces of the Lagrangian read as follows:\footnote{We have ignored the bilinear RPV couplings in this work.}
\begin{align}
{\cal L}_{LLE} =  & \frac{1}{2}\lambda_{ijk}\big[ \widetilde{\nu}_{iL} \bar{e}_{kR} e_{jL} +\widetilde{e}_{jL} \bar{e}_{kR}\nu_{iL} +\widetilde{e}_{kR}^{*} \bar{\nu}_{iL}^c e_{jL} \nonumber \\
& \qquad - (i\leftrightarrow j) \big]+{\rm H.c.}
\label{Eq.RPVLLE} \\
{\cal L}_{LQD}  =   & \lambda^\prime_{ijk}\big[\widetilde{\nu}_{iL}\bar{d}_{kR}d_{jL}+\widetilde{d}_{jL}\bar{d}_{kR}\nu_{iL}+\widetilde{d}^*_{kR}\bar{\nu}^c_{iL}d_{jL}\nonumber \\
&  - \widetilde{e}_{iL}\bar{d}_{kR}u_{jL}-\widetilde{u}_{jL}\bar{d}_{kR}e_{iL}-\widetilde{d}^*_{kR}\bar{e}^c_{iL}u_{jL}\big]+{\rm H.c.}
\label{Eq.RPVLQD}
\end{align}
\begin{figure*}[ht!]
		\centering		\includegraphics[width=0.245\linewidth]{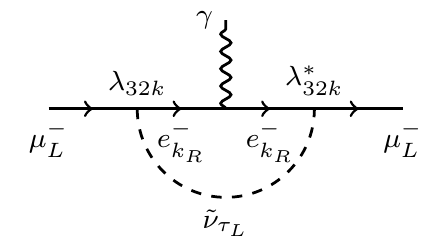}		\includegraphics[width=0.245\linewidth]{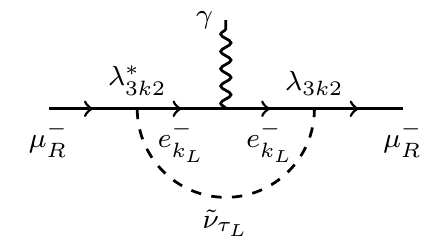}		\includegraphics[width=0.245\linewidth]{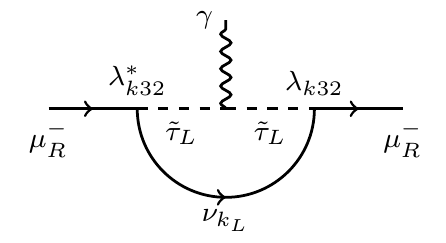} 		 \includegraphics[width=0.245\linewidth]{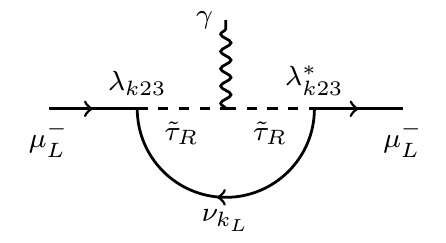} \\
		\includegraphics[width=0.245\linewidth]{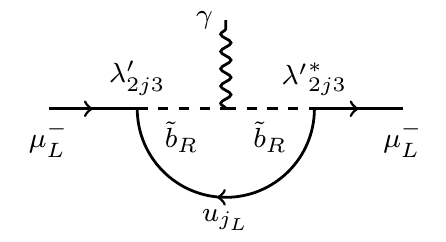}  
		\includegraphics[width=0.245\linewidth]{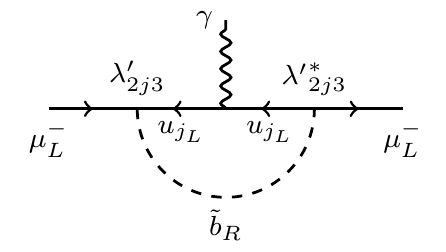}  
		\includegraphics[width=0.245\linewidth]{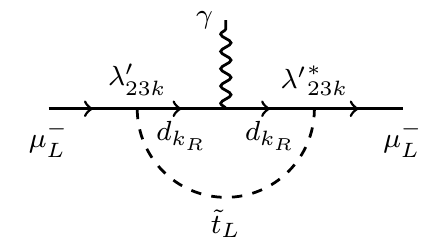} 
  \includegraphics[width=0.245\linewidth]{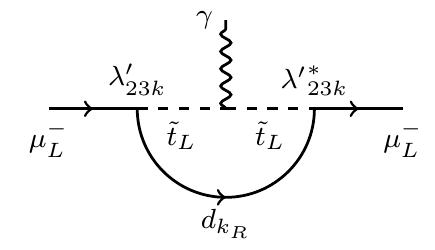}  
	\caption{Relevant contributions to the muon $(g-2)$ from $\lambda$ (top row) and $\lambda'$ (bottom row) couplings in our RPV3 scenario. Note that the stop contributions (the last two diagrams) add up to zero.
	}
	\label{fig:muongm2}
\end{figure*}
Note that the simultaneous presence of $\lambda$ and $\lambda'$ couplings is consistent with proton decay constraints, as long as the relevant $\lambda"$ ($UDD$-type) couplings are either switched off or sufficiently small. 
In general, the above Lagrangians feature $3^3=27$ independent $\lambda'_{ijk}$ couplings and $3^2=9$ independent $\lambda_{ijk}$ couplings.\footnote{$\lambda_{ijk}$ is antisymmetric in the first two indices.} However, in the RPV3 framework with the first two generations of sfermions decoupled, the total number of relevant RPV couplings reduces to $9+6=15$. Moreover, because of the orthogonality between the $R_{D^{(*)}}$, $R_{K^{(*)}}$-preferred region which is mostly controlled by the $\lambda'$ couplings and the muon $(g-2)$-preferred region which is controlled by the  $\lambda$ couplings, we only focus on the $\lambda$ couplings in this paper and find a new solution for the muon $(g-2)$ anomaly with $\lambda_{233}\neq 0$, without affecting the allowed parameter space for $R_{D^{(*)}}$ and $R_{K^{(*)}}$ reported in Ref.~\cite{Dev:2021ipu}.

The RPV3 contributions to $(g-2)_{\mu}$ can, in principle, arise from both $\lambda$ and $\lambda^\prime$ couplings~\cite{Kim:2001se}, as shown in Fig.~\ref{fig:muongm2}. Applying the general results from 
Ref.~\cite{Leveille:1977rc}, one obtains~\cite{Kim:2001se}
\begin{align}
    \Delta a_{\mu} & \ = \ \frac{m_{\mu}^2}{96\pi^2}\sum_{k=1}^3 \left(\frac{2(|\lambda_{32k}|^2+|\lambda_{3k2}|^2)}{m^2_{\widetilde{\nu}_{\tau}}}\right.\nonumber \\
    & \qquad \left.-\frac{|\lambda_{k32}|^2}{m^2_{\widetilde{\tau}_{L}}}
    -\frac{|\lambda_{k23}|^2}{m^2_{\widetilde{\tau}_{R}}}
    +\frac{3|\lambda^\prime_{2k3}|^2}{m_{\widetilde{b}_R}^2} \right)
    \label{eq:gm2l} \, .
\end{align}
Note that the $\lambda'$-contribution, as well as the $\lambda$-contribution from sneutrinos, is always positive definite, whereas the $\lambda$-contribution from staus has the wrong sign and is required to be sub-dominant in order to explain the observed discrepancy in Eq.~\eqref{eq:deltamu}.\footnote{$R$-parity preserving SUSY contributions involving smuons and muon sneutrinos~\cite{Moroi:1995yh,Baum:2021qzx,Chakraborti:2022vds} are small in RPV3 because the first two generations of sfermions are heavy.} 

As shown in Ref.~\cite{Dev:2021ipu}, the $\lambda'$-contribution from sbottom is sub-dominant to the $\lambda$-contribution from sneutrinos, mainly because the LHC lower limits on the masses of colored sfermions like the sbottom are much stronger than those on sneutrinos. In particular, $m_{\widetilde{b}_R}$ is typically between 1.5~TeV and 10~TeV,  and $|\lambda^\prime|/(m_{\widetilde{b}_R}/1\mathrm{\ TeV}) \lesssim 1$ to explain the $R_{D^{(*)}}$ and $R_{K^{(*)}}$ anomalies~\cite{Dev:2021ipu}. This  makes the sbottom contribution to muon $(g-2)$ negligible.

Thus, focusing only on the $\lambda$-contributions in Eq.~\eqref{eq:gm2l}, we see that there are only four relevant couplings, namely,  $\lambda_{132}$, $\lambda_{231}$, $\lambda_{232}$ and $\lambda_{233}$, that lead to a positive contribution to $\Delta a_\mu$ in Eq.~\eqref{eq:gm2l}. However, any two of them cannot be large simultaneously because of the lepton flavor violation constraints from low-energy processes like $\tau^- \to e^- \mu^+ \mu^-$,  $\tau^- \to \mu^-\mu^+\mu^-$, $\mu \to e \gamma$, etc. Therefore, it is safe to assume only one of these couplings to be large, while the rest can be set to zero. 
In Ref.~\cite{Dev:2021ipu}, the nonzero coupling was chosen to be $\lambda_{232}$, which led to four-muon final states at the LHC via resonant sneutrino-pair production. In this paper, we study the case where $\lambda_{233}\neq 0$, which leads to a final state of two muons and two taus at the LHC. We expect this case to be more promising, because of the relatively weaker LHC constraints on signals with tau final states, which in turn are expected to give a weaker bound on the sneutrino mass, thus allowing for a larger contribution to $\Delta a_\mu$, since it is inversely proportional to the square of sneutrino mass [cf.~Eq.~\eqref{eq:gm2l}]. For instance, for $m_{\widetilde \nu_\tau}\sim 100~{\rm GeV}$, $\lambda_{233}\sim 1$ can explain the central value of $\Delta a_\mu$ in Eq.~\eqref{eq:deltamu}.  The remaining two cases, namely with either $\lambda_{132}$ or $\lambda_{231}$ nonzero, which give rise to final states with two electrons and two muons, will give a bound on the sneutrino mass comparable to that in the four-muon case studied in Ref.~\cite{Dev:2021ipu}.  

We have also assumed $\lambda'_{311}$ to be small in order to avoid the resonance production of $\widetilde\nu_\tau$, which gives stringent bounds from the LHC. For $\lambda^\prime_{311}=0.1$, the limit on $m_{\widetilde{\nu}_{\tau}}$ is $\mathcal{O} ({\rm TeV})$~\cite{ATLAS:2018mrn}. For the sub-TeV $\widetilde{\nu_{\tau}}$ considered here, we therefore need $\lambda^\prime_{311} < \mathcal{O} (0.01)$.

\subsection{Low-energy Constraints} 
With $\lambda_{233}=-\lambda_{323}\neq 0$ (and all other $\lambda_{ijk}=0$), the left-handed stau contribution to $\Delta a_\mu$ in Eq.~\eqref{eq:gm2l} is absent. As for the right-handed stau contribution, which is of the wrong sign, we need to make sure that it is sub-dominant to the sneutrino contribution. This is automatically enforced by the low-energy constraint from tau decay, because $\widetilde{\tau}_R$ with coupling $\lambda_{233}\neq 0$ has a tree-level contribution to the process $\tau \to \mu \nu \bar{\nu}$. The effective four-fermion Lagrangian for the tau decay (after integrating out the $\widetilde{\tau}_R$) is
\begin{align}
    {\cal L}^{\lambda_{233}}_{\tau \to \mu \nu \bar{\nu}} \ =  \ -\frac{|\lambda_{233}|^2}{2m^2_{\widetilde{\tau}_R}}(\overline{\mu}_L \gamma^\mu \nu_{\mu L})(\overline{\nu}_{\tau L} \gamma_\mu \tau_L).
\end{align}
The effective Lagrangian has the same chiral structure as the SM contribution to tau decay. This can only affect the $g_{LL}^V$ coupling (in the notation of Ref.~\cite{Fetscher:1986uj}),  and because of the normalization condition of the couplings, our scenario does not influence the Michel parameters~\cite{Kuno:1999jp}.

However, it still affects the $e-\mu$ universality in tau decays, measured by the ratio 
\begin{align}
    R_{\mu e} \equiv \frac{\Gamma(\tau \to \mu\nu\bar{\nu}) }{ \Gamma(\tau \to e\nu\bar{\nu})} \, .
\end{align}
The SM prediction including mass effects gives $R_{\mu e}^{\rm SM} = 97.26\%$ while the experimental measurement prefers a slightly larger central value $R_{\mu e}^{\rm exp} = (97.62 \pm 0.28)\%$~\cite{Workman:2022ynf}. The ratio between the experimental value and the theory prediction in our scenario is given by 
\begin{align}
    \frac{R_{\mu e}^{\rm exp}}{R_{\mu e}^{\rm SM}} \simeq \left( 1 + \frac{1}{4\sqrt{2}G_{\rm F}} \frac{|\lambda_{233}|^2}{m_{\widetilde{\tau}_R}^2} \right)^2 \, ,
    \label{eq:universal}
\end{align}
where $G_F$ is the usual Fermi constant. Allowing for $3\sigma$ uncertainty in the experimental value, we obtain a limit on $\lambda_{233}$ as
\begin{align}
    |\lambda_{233}| \lesssim 0.65\left( \frac{m_{\widetilde{\tau}_R}}{1 {\rm \ TeV}} \right).
\end{align}

A slightly stronger limit can be derived by comparing the decays $\tau \to \mu \nu \bar{\nu}$ and $\mu \to e \nu \bar{\nu}$~\cite{Barger:1989rk}, which is described by the observable 
\begin{align}
    R_{\tau / \mu} \equiv \frac{{\rm BR}(\tau \to \mu \nu \bar{\nu})_{\rm exp} / {\rm BR}(\tau \to \mu \nu \bar{\nu})_{\rm SM}}{{\rm BR}(\mu \to e \nu \bar{\nu})_{\rm exp} / {\rm BR}(\mu \to e \nu \bar{\nu})_{\rm SM}} .
\end{align}
The current value is measured to be $R_{\tau / \mu} = 1.0022 \pm 0.0030$~\cite{Trifinopoulos:2018rna}. Using expressions analogous to Eq.~\eqref{eq:universal}, and taking $3\sigma$ uncertainties in the measured value, it converts to a slightly stronger bound on $\lambda_{233}$:
\begin{align}
    |\lambda_{233}| \lesssim 0.61\left( \frac{m_{\widetilde{\tau}_R}}{1 {\rm \ TeV}} \right).
\label{eq:tau decay}
\end{align}
Eq.~\eqref{eq:tau decay} is satisfied for any $|\lambda_{233}| < \sqrt{4\pi}$ (perturbative limit), as long as $m_{\widetilde{\tau}_R} \gtrsim 5.8$ TeV. For such $m_{\widetilde{\tau}_R}$ values, the $\widetilde{\tau}_R$ contribution to $\Delta a_\mu$ can be safely neglected.


\subsection{Neutrino Mass Constraint} 
The $LLE$ interactions contribute to neutrino mass at one-loop level through the lepton-slepton loop~\cite{Hall:1983id,Babu:1989px,Davidson:2000ne, Barbier:2004ez}. In the RPV3 scenario, we have 
\begin{align}
    M_{ij}^\nu \simeq & \frac{1}{16\pi^2}\sum_k \lambda_{ik3}\lambda_{j3k}m_{e_k}\frac{\left(\widetilde{m}_{LR}^e\right)^2_{33}}{m^2_{\widetilde \tau_R}-m^2_{\widetilde \tau_L}}\log\left(\frac{m^2_{\widetilde \tau_R}}{m^2_{\widetilde \tau_L}} \right) \nonumber \\
    & \quad +(i\leftrightarrow j) \, ,
    \label{eq:neutrino}
\end{align}
where $\left(\widetilde{m}_{LR}^e\right)^2$ is the left-right slepton mixing matrix, given by 
\begin{align}
    \left(\widetilde{m}_{LR}^e\right)^2_{ij} = \frac{v_d}{\sqrt 2}\left(A^e_{ij}-\mu \tan\beta y_{ij}^e\right) \, ,
\end{align}
where $A^e$ is the soft trilinear term, $\mu$ is the Higgs-Higgs mixing (or off-diagonal Higgsino mass) term, $y^e$ is the lepton Yukawa coupling, and $\tan\beta=v_u/v_d$ is the ratio of the vacuum expectation values of the two Higgs doublets.  In the basis of diagonal charged lepton masses, it is customary to assume that the $A$-term is proportional to the Yukawa coupling, {\it i.e.} $A^e_{33}=A^\tau y^\tau$. We also assume that $m_{\widetilde{\tau}_L} = m_{\widetilde{\tau}_R}$, in which case $\log\left(m^2_{\widetilde \tau_R}/m^2_{\widetilde \tau_L} \right)/\left(m^2_{\widetilde \tau_R}-m^2_{\widetilde \tau_L}\right)=1/m^2_{\widetilde \tau_R}$. Then Eq.~\eqref{eq:neutrino} simplifies to 
\begin{align}
    M_{23}^\nu & \simeq \frac{|\lambda_{233}|^2}{8\pi^2}\frac{m_\tau^2}{m^2_{\widetilde \tau_R}}(A^\tau-\mu \tan\beta) \nonumber \\
    & = (0.05~{\rm eV})|\lambda_{233}|^2
    \left(\frac{6~{\rm TeV}}{m_{\widetilde\tau_R}}\right)^2  \frac{(A^\tau-\mu \tan\beta)}{45~{\rm MeV}} \, .
\end{align}
Thus the neutrino mass constraint can be easily satisfied, albeit with some fine-tuning in the SUSY parameters $A^\tau$ and $\mu\tan\beta$, which however do not affect the muon $(g-2)$ solution in our case.  

\section{Signal and background analysis}
\label{sec:collider}
We use the results of the analysis done by the ATLAS collaboration in Ref.~\cite{ATLAS:2021yyr}, with the data recorded during Run-2 of the LHC at a center-of-mass energy of $\sqrt s = 13$~TeV and integrated luminosity of 139~fb$^{-1}$, which targeted a search for RPV-SUSY in final states with four or more charged leptons (electrons, muons and taus). Later we will emphasize how the signal sensitivity can be enhanced with more dedicated selections. 

\begin{figure}[t!]
    \centering
    \includegraphics[width=0.35\textwidth]{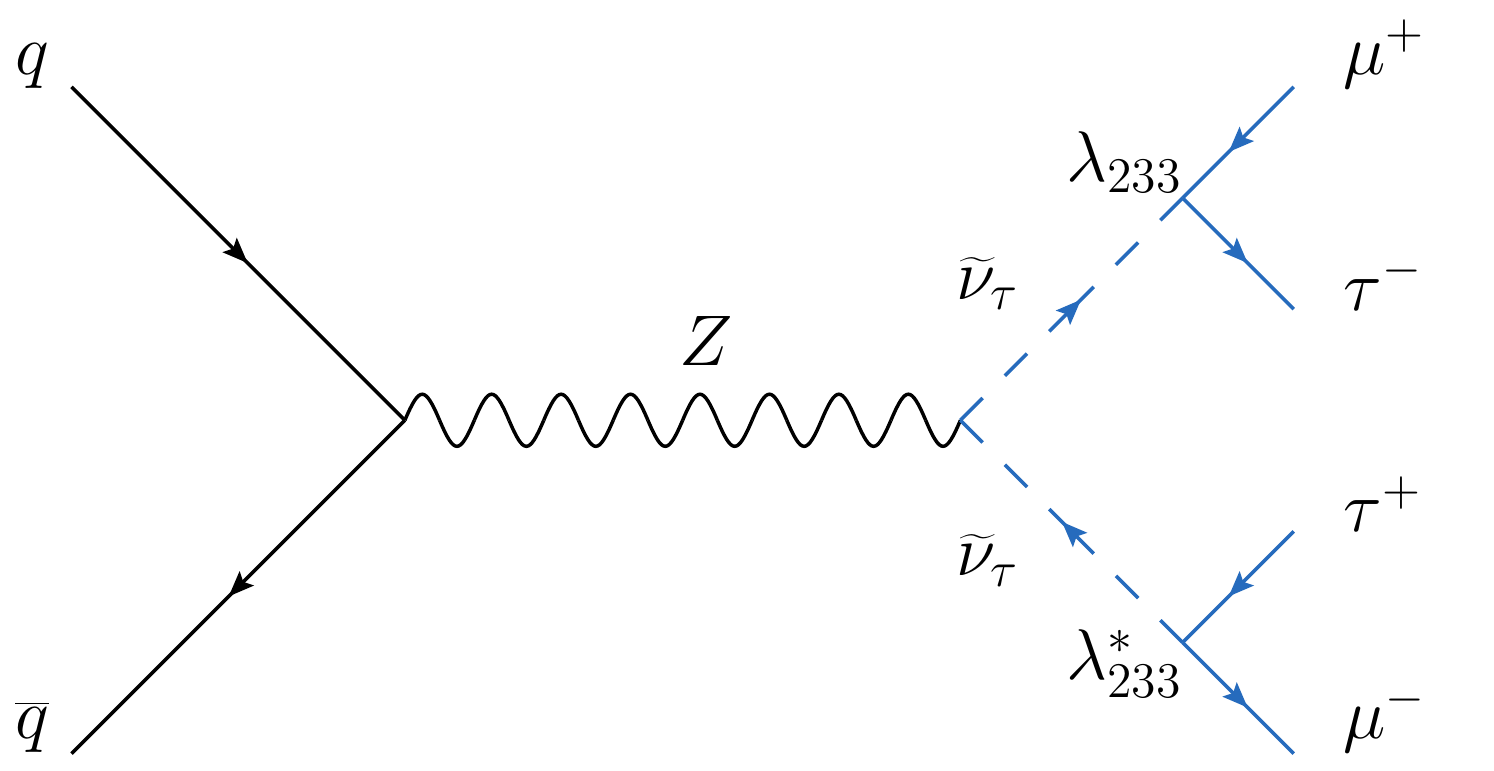}
    \caption{Feynman diagram for the $\tau^+\tau^-\mu^+\mu^-$ signal from the sneutrino pair-production in our RPV3 scenario. The blue portion of the figure is closely related to the muon $(g-2)$, {\it i.e.} if we join the $\tau$ legs and attach a photon to it, it resembles the first two diagrams in Fig.~\ref{fig:muongm2}. }
    \label{fig:signal}
\end{figure}

The $\tau^+\tau^-\mu^+\mu^-$ signal that is relevant to the muon $(g-2)$-anomaly comes from the sneutrino pair-production, followed by each sneutrino decaying into
$\tau^-\mu^+$ pair via the $\lambda_{233}$ coupling, as shown in Fig.~\ref{fig:signal}. Note that there are also some contributions to this final state from pair production of muons or taus, followed by sneutrino single production from a lepton leg and its subsequent decay into $\tau^-\mu^+$ pair. However, in the parameter space of interest, we find that the sneutrino single production contributes far less than the pair production shown in Fig.~\ref{fig:signal}. Also note that because of the particular structure of the $LLE$ interaction terms in Eq.~\eqref{Eq.RPVLLE}, $\widetilde{\nu}_\tau \to \tau^+\mu^-$ is not allowed if we assume only $\lambda_{233} \neq 0$, and thus, we cannot have a more distinguishing signal like $\tau^+\tau^+\mu^-\mu^-$ or $\tau^-\tau^-\mu^+\mu^+$ in our 
scenario.

\subsection{Simulated Event Samples}

All event samples for the signal and the SM backgrounds were generated using {\sc MadGraph5\_aMC@NLO}~\cite{Alwall:2014hca} at leading order (LO) parton-level. 
The SM background events are not used directly for the estimation of the sensitivity, but as a cross-check that after applying the selections stated in Ref.~\cite{ATLAS:2021yyr} we get a similar background estimation. 
In addition, we use the simulation of the SM backgrounds in order to estimate the efficiency of our proposed dedicated selection.
For the RPV-SUSY signal, a dedicated universal FeynRules Output (UFO) model was produced using {\sc FeynRules}~\cite{Alloul:2013bka}.
For all of the samples, both signal and background, the 5-flavor scheme was used for the event generation with the NNPDF30LO parton distribution function (PDF) set~\cite{NNPDF:2014otw} and the default {\sc MadGraph5\_aMC@NLO} LO dynamical scale, which is the transverse mass calculated by a $k_t$-clustering of the final-state partons~\cite{Catani:1993hr}.
After generation, events were interfaced with the {\sc Pythia~8}~\cite{Mrenna:2016sih} parton shower, and different jet-multiplicities were matched using the MLM scheme~\cite{Mangano:2006rw} with the default {\sc MadGraph5\_aMC@NLO} parameters. 
Finally, all samples were processed through {\sc Delphes 3}~\cite{deFavereau:2013fsa}, which simulates the detector effects, applies simplified reconstruction algorithms and was used for the reconstruction of all objects.

According to Ref.~\cite{ATLAS:2021yyr}, the dominant SM backgrounds are $ZZ$, $t \bar t Z$, $VVV$ ($V = W, Z$) and Higgs production. 
We note that off-shell production is included for $W$ and $Z$.
All of those processes can have four leptons in the final state, similar to our signal. 
All of those backgrounds were simulated and similar selections of the analysis as in Ref.~\cite{ATLAS:2021yyr} were applied.
In addition to those irreducible backgrounds, there are dominant reducible backgrounds that contain processes that have at least one fake lepton, such as $t \bar t$, $Z+$jets, $WZ$, $WW$, $WWW$, $t \bar t W$.

\subsection{Event selection and background}
The reconstruction of electrons and muons (light leptons) was done based on efficiency parametrization which depends on transverse momentum ($p_{\mathrm{T}}$) and pseudo-rapidity ($\eta$), and with an isolation from other energy-flow objects applied in a cone of $\Delta R=0.4$.
Electrons must have $|\eta| < 2.47$~GeV and $p_{\mathrm{T}} > 7$~GeV, while muons are required to have $|\eta| < 2.7$~GeV and $p_{\mathrm{T}} > 5$~GeV.

The reconstruction of jets was done using the anti-$k_{t}$~\cite{Cacciari:2008gp} clustering algorithm with radius parameter of $R=0.4$ implemented in {\sc FastJet}~\cite{Cacciari:2011ma,Cacciari:2005hq}.
Jets are required to have $p_{\mathrm{T} } >20$~GeV and $\left|\eta\right|<2.8$.
The identification of $b$-tagged jets was done by applying a $p_{\mathrm{T}}$-dependent weight based on the jet’s associated flavor and the MV2c20 tagging algorithm~\cite{ATL-PHYS-PUB-2015-022} in the 70\% working point, which is the default one provided by {\sc Delphes 3}.\footnote{We note that in Ref.~\cite{ATLAS:2021yyr} the 85\% working point is used for $b$-tagging, but since we have no $b$-jets in our signal production, the impact of this difference on the signal selection is negligible.}

Hadronically decaying taus have a visible part coming from the hadrons involved in the process and an invisible part coming from the neutrino. 
The visible part ($\tau_\mathrm{had}^\mathrm{vis}$) is reconstructed using jets, with $|\eta| < 1.37$ or $1.52< |\eta| < 2.47$ and $p_{\mathrm{T} } > 20$~GeV~\cite{ATL-PHYS-PUB-2015-045}, using information about the tracks within $\Delta R = 0.2$ of the jet direction.

The missing transverse momentum $\vec p_{\mathrm{T}}^\mathrm{miss}$ and its magnitude $E_{\mathrm{T}}^\mathrm{miss}$ are reconstructed as the negative sum of the $p_\mathrm{T}$ of all objects in the event and a soft term built from all tracks not associated to any reconstructed object.

\begin{figure}
    \centering
    \includegraphics[width=0.49\textwidth]{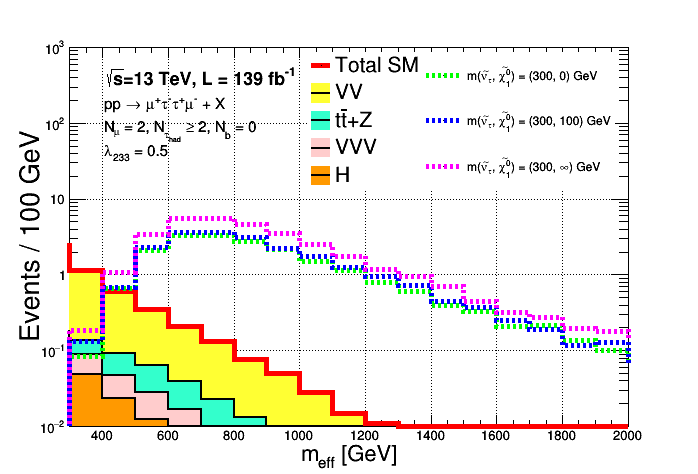}
    \caption{Distribution of the effective mass $m_\mathrm{eff}$ defined in Eq.~\eqref{eq:meff}. All of the selections of $\text{SR2}_{\text{bveto}}^{\text{loose}}$-$\mu\mu$ and $\text{SR2}_{\text{bveto}}^{\text{tight}}$-$\mu\mu$ are implemented, as described in Tab.~\ref{tab:selections}, beside the $m_\mathrm{eff}$ selection. Three signal points are presented by setting $m_{\widetilde{\nu}_\tau}=300$~GeV
    and $\lambda_{233}=0.5$, for three benchmark values of $m_{\widetilde{\chi}_1^0}$: with a very small value (0~GeV), 100~GeV, and a very large value ($\infty$). }
    \label{fig:meff}
\end{figure}

The event selection applied in Ref.~\cite{ATLAS:2021yyr} which yields the best sensitivity for the signal scenario considered here, is noted with two Signal Regions (SRs): $\text{SR2}_{\text{bveto}}^{\text{loose}}$ and $\text{SR2}_{\text{bveto}}^{\text{tight}}$. These SRs contain two light leptons (electrons or muons) and at least two $\tau_\mathrm{had}^\mathrm{vis}$.
In addition, a $b$-veto is applied by requiring no $b$-tagged jets in the events.
In order to reduce events with a $Z$-boson decaying to a pair of leptons, events with a pair of opposite-sign and same-flavor (OSSF) leptons within the mass range of $81.2-111.2$~GeV are removed. 
The main discriminating variable used in Ref.~\cite{ATLAS:2021yyr} is $m_\mathrm{eff}$, defined as:
\begin{equation}
    m_\mathrm{eff} = \sum_i p_{\mathrm{T},\ell_i} + \sum_j p_{\mathrm{T},\tau_{\mathrm{had}_j}^\mathrm{vis}} + \sum_k p_{\mathrm{T},\mathrm{jet}_k > 40} + E_\mathrm{T}^\mathrm{miss} ,
    \label{eq:meff}
\end{equation}
where $p_{\mathrm{T},\ell_i}$ is the $p_\mathrm{T}$ of a light lepton, $p_{\mathrm{T},\tau_{\mathrm{had}_i}^\mathrm{vis}}$ is the $p_\mathrm{T}$ of a $\tau_\mathrm{had}^\mathrm{vis}$ and $p_{\mathrm{T},\mathrm{jet}_i > 40}$ is the $p_\mathrm{T}$ of a jet with a minimum transverse momentum of 40~GeV.
$\text{SR2}_{\text{bveto}}^{\text{loose}}$ and $\text{SR2}_{\text{bveto}}^{\text{tight}}$ differ from each other by a looser or a tighter selection of $m_\mathrm{eff}$, respectively.
Based on these SRs, we emphasize how the results would improve with a dedicated selection of only two muons as the light leptons. We call these selections $\text{SR2}_{\text{bveto}}^{\text{loose}}$-$\mu\mu$ and $\text{SR2}_{\text{bveto}}^{\text{tight}}$-$\mu\mu$.
All of the selections are summarized in Table \ref{tab:selections}.
The distribution of $m_\mathrm{eff}$ with a selection of only two muons as the light leptons is shown in Fig.~\ref{fig:meff}, along with the dominant backgrounds. Additional distributions with our improved selection are given in Appendix~\ref{sec:kinematic}.

\begin{table}[b!]
\caption{Selections for the analysis. $\text{SR2}_{\text{bveto}}^{\text{loose}}$ and $\text{SR2}_{\text{bveto}}^{\text{tight}}$ apply the selection used in Ref.~\cite{ATLAS:2021yyr}, while $\text{SR2}_{\text{bveto}}^{\text{loose}}$-$\mu\mu$ and $\text{SR2}_{\text{bveto}}^{\text{tight}}$-$\mu\mu$ apply similar selections, but with only muons as the light leptons.}
\label{tab:selections}
\resizebox{\hsize}{!}{
\begin{tabular}{l c c c c}
\tabularnewline
\hline 
Selection & $\text{SR2}_{\text{bveto}}^{\text{loose}}$ & $\text{SR2}_{\text{bveto}}^{\text{loose}}$-$\mu\mu$ & $\text{SR2}_{\text{bveto}}^{\text{tight}}$ & $\text{SR2}_{\text{bveto}}^{\text{tight}}$-$\mu\mu$ \\
\hline 
$N_\ell$ & \multicolumn{4}{c}{$=2$}
\tabularnewline
$N_\mu$ & 0-2 & $=2$ & 0-2 & $=2$
\tabularnewline
$N_e$ & 0-2 & $=0$ & 0-2 & $=0$
\tabularnewline
$N_{\tau_\mathrm{had}^\mathrm{vis}}$ &  \multicolumn{4}{c}{$\geq 2$}
\tabularnewline
$N_{b}$ &  \multicolumn{4}{c}{$= 0$}
\tabularnewline
$m_{\ell \ell}^\mathrm{OSSF}$~[GeV] & \multicolumn{4}{c}{$<81.2 \And >101.2$}
\tabularnewline
$m_\mathrm{eff}$~[GeV] & \multicolumn{2}{c}{$>600$} & \multicolumn{2}{c}{$>1000$} 
\tabularnewline
\hline 
\end{tabular}
}
\end{table}

\subsection{Bounds from Current Data}


The observed number of signal events with 95\% confidence level (CL), $S^{95}_\mathrm{obs}$, is reported in Ref.~\cite{ATLAS:2021yyr}.
The meaning of this number is that given a signal hypothesis, if the expected yield in the signal region is higher than $S^{95}_\mathrm{obs}$, the signal hypothesis is excluded with 95\% CL.
For the selection described above, these values are 8.45 and 5.63 for  $\text{SR2}_{\text{bveto}}^{\text{loose}}$ and $\text{SR2}_{\text{bveto}}^{\text{tight}}$, respectively.
Using these numbers, we set limits on our signal hypothesis.

\subsection{Expected Improved Bounds}
Given the limits set by using an existing analysis, a few remarks are in place:
\begin{itemize}
    \item In Ref.~\cite{ATLAS:2021yyr} an inclusive selection of the light lepton flavor is done. In the signal hypothesis mentioned in this paper, only final states with two muons are relevant. This selection is expected to reduce the SM irreducible background, and has no impact on the signal scenario we consider.
    \item About a third of the background contribution in $\text{SR2}_{\text{bveto}}^{\text{loose}}$-$\mu\mu$ is coming from the reducible background. Typically, this background is more dominant for final states with electrons. Therefore, excluding events with electrons is expected to remove a significant part of the reducible background.
\end{itemize}

In order to estimate how would $S^{95}_\mathrm{obs}$ change given our new selection, we calculate the expected $Z$-value, which is the number of standard deviations from the background-only hypothesis given a signal yield and background uncertainty, using the \verb|BinomialExpZ| function by \verb|RooFit|~\cite{Verkerke:2003ir}.
We scan over different values of the signal yield. Once we get similar $Z$-value to the ones from $\text{SR2}_{\text{bveto}}^{\text{loose}}$ and $\text{SR2}_{\text{bveto}}^{\text{tight}}$ in Ref.~\cite{ATLAS:2021yyr}, we set $S^{95}_\mathrm{obs}$ of $\text{SR2}_{\text{bveto}}^{\text{loose}}$-$\mu\mu$ and $\text{SR2}_{\text{bveto}}^{\text{tight}}$-$\mu\mu$.
We do the same procedure for two values of the total integrated luminosity: $139.0~\mathrm{fb}^{-1}$, as in Ref.~\cite{ATLAS:2021yyr}, which corresponds to the total integrated luminosity recorded during Run-2 of the LHC, and $3000.0~\mathrm{fb}^{-1}$, which corresponds to the expected integrated luminosity from the HL-LHC.

\section{Results} \label{sec:results}
\begin{figure*}[t!]
    \centering
    \subfloat[$m_{\widetilde{\chi}_1^0}\ll m_{\widetilde{\nu}_\tau}$]{\includegraphics[width=0.495\textwidth]{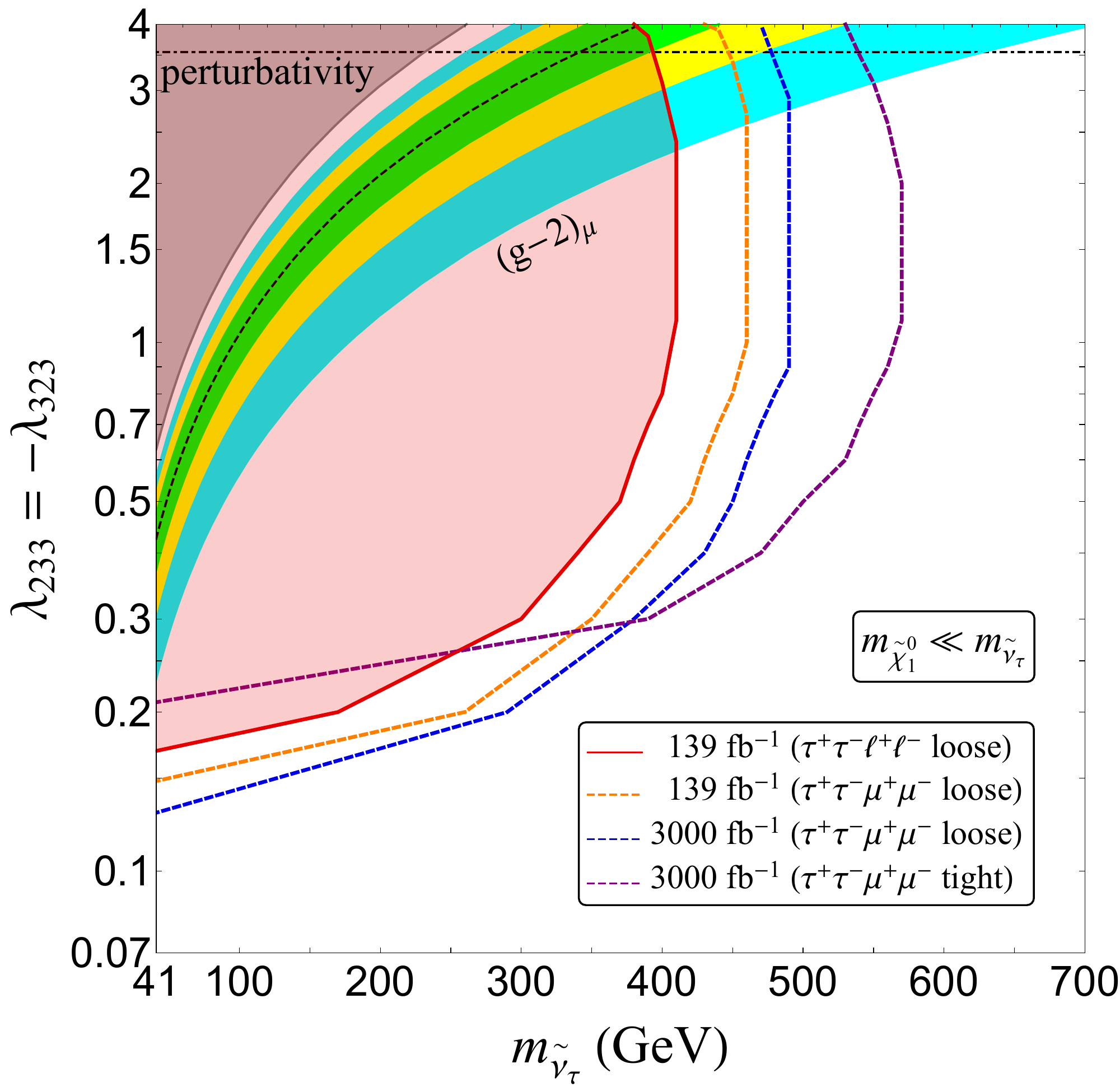}}
    \subfloat[$m_{\widetilde{\chi}_1^0}=100$ GeV]{\includegraphics[width=0.495\textwidth]{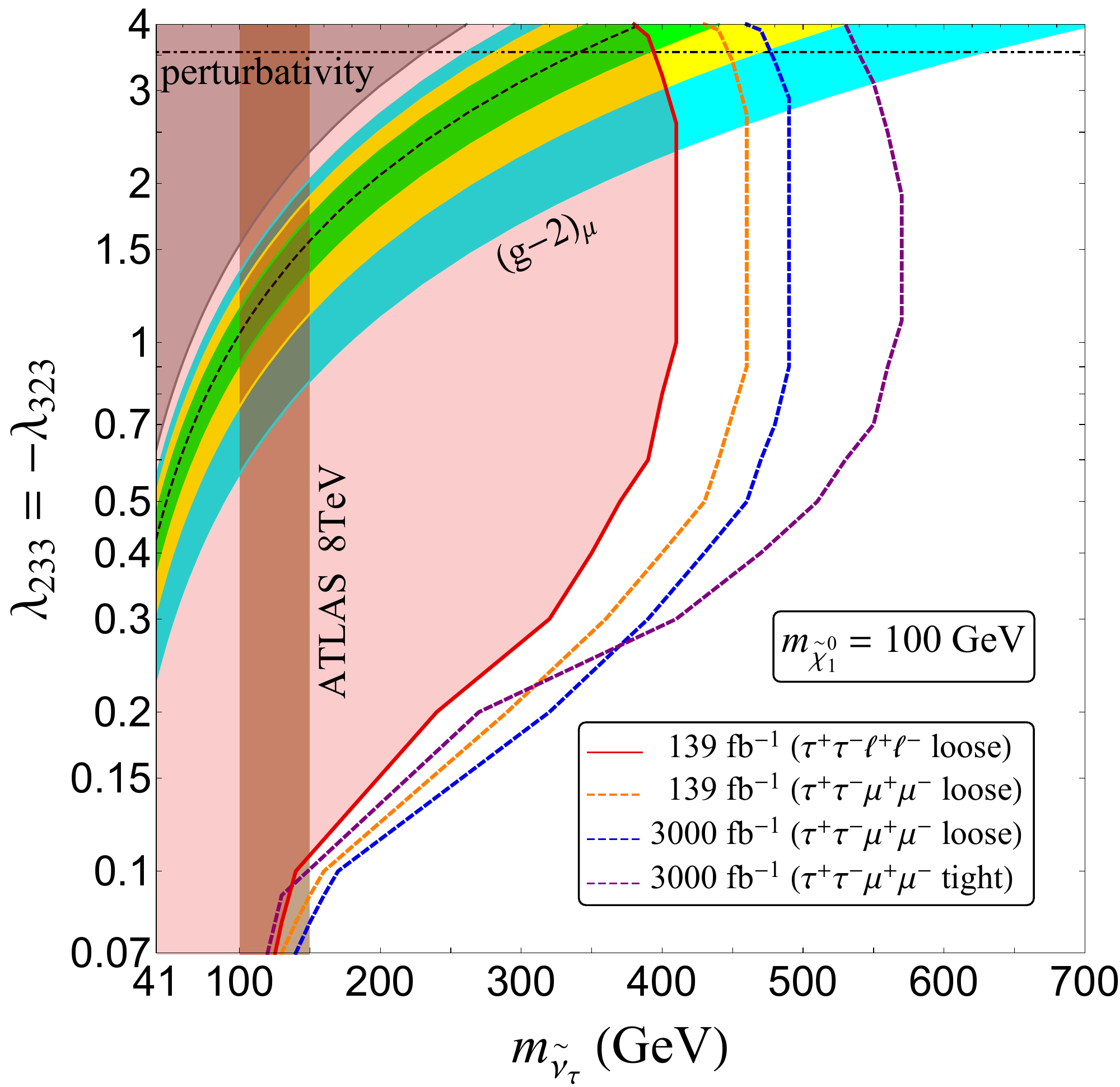}}\\
    \subfloat[$m_{\widetilde{\chi}_1^0}\gg m_{\widetilde{\nu}_\tau}$]{\includegraphics[width=0.495\textwidth]{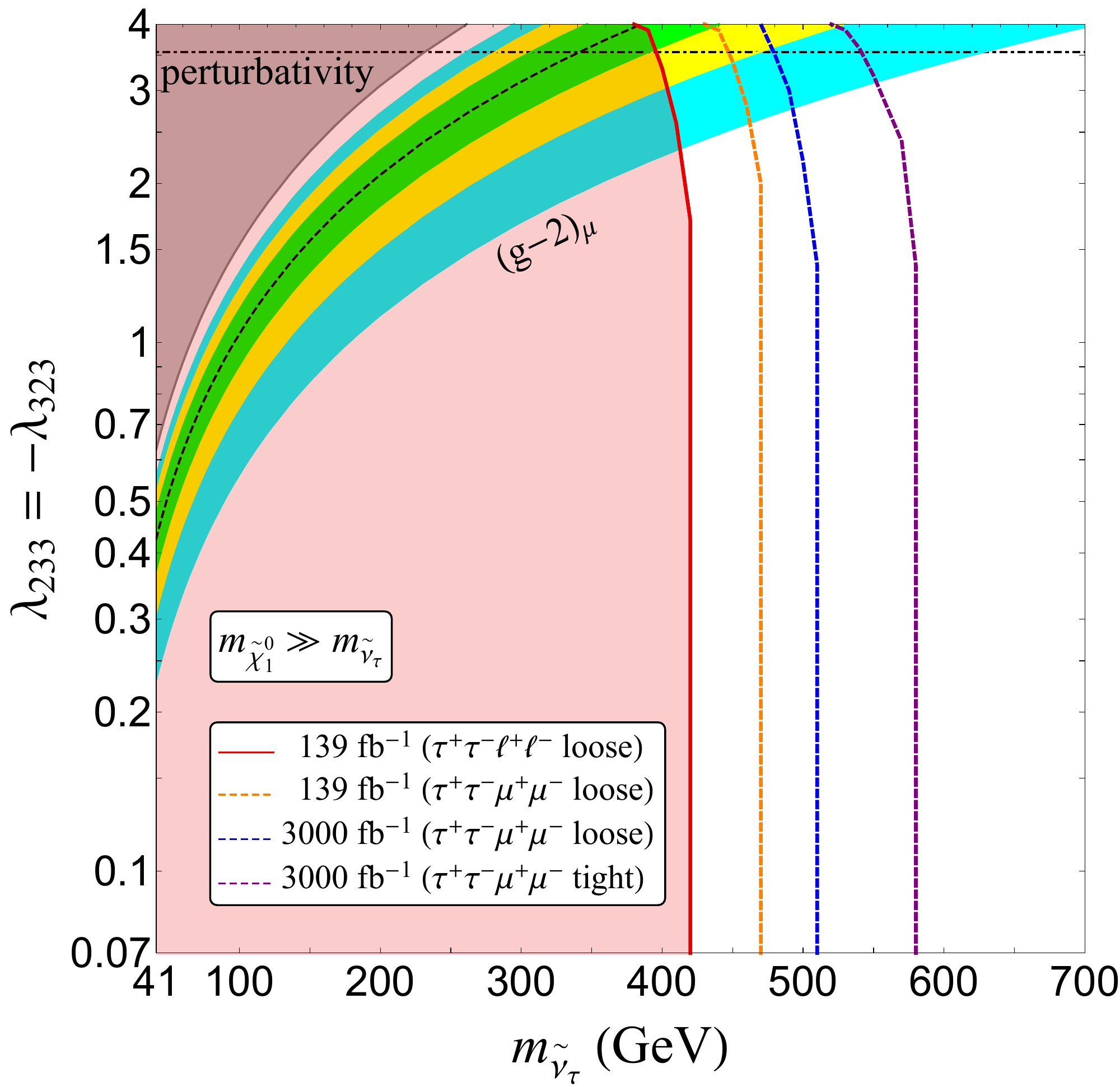}}
    \caption{Three benchmark cases of our RPV3 scenario with $m_{\widetilde{\chi}_1^0}$ (a) much smaller than $m_{\widetilde{\nu}_\tau}$, (b) 100 GeV, and (c) much larger than $m_{\widetilde{\nu}_\tau}$)  in the $(m_{\widetilde{\nu}_\tau},\lambda_{233})$ parameter space. The red (solid) and orange (dashed) contours, respectively, show the 95\% CL current bounds derived from the $139~\mathrm{fb}^{-1}$ LHC data in the $\tau^+\tau^-\ell^+\ell^-$ $\text{SR2}_{\text{bveto}}^{\text{loose}}$ channel, and the expected improved bounds with the same dataset in the $\tau^+\tau^-\mu^+\mu^-$ $\text{SR2}_{\text{bveto}}^{\text{loose}}$-$\mu\mu$ channel, whereas the blue and purple (dashed) contours show the 95\% CL sensitivities at HL-LHC with  $3000~\mathrm{fb}^{-1}$ luminosity in the $\tau^+\tau^-\mu^+\mu^-$ $\text{SR2}_{\text{bveto}}^{\text{loose}}$-$\mu\mu$ and $\text{SR2}_{\text{bveto}}^{\text{tight}}$-$\mu\mu$ channels, respectively.  The green, yellow and cyan-shaded regions explain the muon $(g-2)$-anomaly at $1\sigma$, $2\sigma$ and $3\sigma$, respectively, while the black solid curve at the middle of the green region gives the best-fit value. The gray-shaded region on the top left corner is the $5\sigma$-exclusion region from muon $(g-2)$. The brown-shaded region in case (b) is excluded by an 8 TeV LHC multi-lepton search~\cite{ATLAS:2014pjz} [not applicable to cases (a) and (c)]. The horizontal black dot-dashed line shows the perturbativity limit.}
    \label{fig:results}
\end{figure*}

We consider three benchmark cases of our scenario where the mass of the lightest neutralino $m_{\widetilde{\chi}_1^0}$ is (a) much smaller than $m_{\widetilde{\nu}_\tau}$, (b) equal to a fixed value of 100 GeV, and (c) much larger than $m_{\widetilde{\nu}_\tau}$. In Fig.~\ref{fig:results}, the red solid contours show our 95\% CL bounds derived in each case  in the $(m_{\widetilde{\nu}_\tau},\lambda_{233})$ parameter space from the current 13~TeV LHC Run-2 data with $139~\mathrm{fb}^{-1}$ in the $\tau^+\tau^-\ell^+\ell^-$ channel with 
$\text{SR2}_{\text{bveto}}^{\text{loose}}$ selection, as given in Tab.~\ref{tab:selections}. The orange dashed contours, on the other hand, show the expected improved bounds derived from the same LHC dataset in the $\tau^+\tau^-\mu^+\mu^-$ channel with 
$\text{SR2}_{\text{bveto}}^{\text{loose}}$-$\mu\mu$ selection, {\it i.e.} excluding the electron final states from the selection. The blue and purple dashed contours are the expected improved bounds from the HL-LHC with $3000~\mathrm{fb}^{-1}$ luminosity in the  $\tau^+\tau^-\mu^+\mu^-$ channel with 
$\text{SR2}_{\text{bveto}}^{\text{loose}}$-$\mu\mu$ and with 
$\text{SR2}_{\text{bveto}}^{\text{tight}}$-$\mu\mu$ selections, respectively. 

We do not show the $\text{SR2}_{\text{bveto}}^{\text{tight}}$ selection results for the $139~\mathrm{fb}^{-1}$ case, because they are found to be weaker than the corresponding  $\text{SR2}_{\text{bveto}}^{\text{loose}}$ results. However, as shown in Fig.~\ref{fig:results}, this is not the case for the 3000 $\mathrm{fb}^{-1}$ luminosity, where the tight selection gives better results than the loose selection in the large coupling region. 

Our analysis for the existing 139 fb$^{-1}$ LHC data uses the selection of $m_{\rm eff} > 600$~GeV (``loose'') while our HL-LHC analysis also uses $m_{\rm eff} > 1000$~GeV (``tight''). It turns out that when the mass of sneutrino is relatively small, the leptons in the final state are too soft to pass the tight selection of $m_{\rm eff} > 1000$~GeV. This feature makes the bounds of HL-LHC (${\rm SR2_{bveto}^{tight}}$-$\mu\mu$) weaker than the 139 fb$^{-1}$ LHC (${\rm SR2_{bveto}^{loose}}$) in the small sneutrino mass region, as can be seen from Fig.~\ref{fig:results} (a) and (b).

Some of the features in Fig.~\ref{fig:results} are the same as those found in the four-muon channel~\cite{Dev:2021ipu}. In particular, the LHC bounds are nearly vertical, with a lower limit on the sneutrino mass of $m_{\widetilde{\nu}_\tau}\gtrsim 400$ GeV, when $\lambda_{233}$ is large or when $m_{\widetilde{\nu}_{\tau}} \ll m_{\widetilde{\chi}_1^0}$ because the dilepton branching ratio of the sneutrino $\mathrm{BR}(\widetilde{\nu}_{\tau} \to \tau^-\mu^+)$ is dominant in these regions. In Fig.~\ref{fig:results} (a) and (b), the bounds slowly bend toward the horizontal direction as we decrease the coupling $\lambda_{233}$ because the $\mathrm{BR}(\widetilde{\nu}_{\tau} \to \widetilde{\chi}_1^0\nu_\tau)$ governed solely by the $R$-parity conserving gauge coupling (and hence, independent of the $\lambda_{233}$ coupling) becomes more and more important. Finally, as the mass of the sneutrino gets close to the mass of the neutralino, the bounds asymptotically approach the line $m_{\widetilde{\nu}_{\tau}} = m_{\widetilde{\chi}_1^0}$ because the $\widetilde{\nu}_{\tau}\to \widetilde{\chi}_1^0\nu_\tau$ decay becomes kinematically suppressed in this region and $\mathrm{BR}(\widetilde{\nu}_{\tau} \to \tau^-\mu^+)$ is dominant again. This asymptotic feature is out of the range in Fig.~\ref{fig:results} (a) as the $m_{\widetilde{\nu}_{\tau}}$ value starts from the model-independent lower limit of 41 GeV,  derived from the LEP data on the invisible $Z$ decay width~\cite{ALEPH:1991qhf}. 

The vertical brown-shaded region in Fig.~\ref{fig:results} (b) (where $m_{\widetilde{\chi}_1^0} = 100$~GeV) is excluded by an old 8 TeV LHC multi-lepton search~\cite{ATLAS:2014pjz}. But for the cases (a) $m_{\widetilde{\chi}_1^0} \ll m_{\widetilde{\nu}_{\tau}}$ and (c) $m_{\widetilde{\chi}_1^0} \gg m_{\widetilde{\nu}_{\tau}}$, this search does not apply because the mass of the lightest neutralino is outside the range of their assumption.

The green, yellow and cyan-shaded regions in Fig.~\ref{fig:results} explain the muon $(g-2)$-anomaly at $1\sigma$, $2\sigma$ and $3\sigma$ CL, respectively, while the black dashed curve gives the best-fit value. The gray-shaded region on the top left corner gives a $\Delta a_\mu$ discrepancy of more than $5\sigma$, and hence, is disfavored. From Fig.~\ref{fig:results}, we see that the new LHC limits derived here preclude most of the muon $(g-2)$-preferred region in our RPV3 scenario, except for large $\lambda_{233}$ coupling values close to the perturbative limit (shown by the horizontal black dot-dashed line). The future HL-LHC projected sensitivities shown here could completely cover the remaining $2\sigma$-preferred regions. It should be noted here that the lower boundaries of the yellow and cyan-shaded regions correspond to corrections of the muon $(g-2)$ at $2\sigma$ (with $\Delta a_\mu = 133 \times 10^{-11}$) and $3\sigma$ (with $\Delta a_\mu = 74 \times 10^{-11}$), respectively. If the new lattice results for the SM prediction come closer to the BMW-reported one, the new central value for $\Delta a_\mu$ is expected to lie somewhere between these two lower boundaries, which in fact opens up a larger allowed parameter space below the perturbativity limit that can be probed at the HL-LHC.

For completeness, we also considered other possible experimental limits for the case (a) $m_{\widetilde{\chi}_1^0} \ll m_{\widetilde{\nu}_{\tau}}$ that could potentially be relevant to the parameter space considered here. In particular, we analyzed the LHC mono-jet~\cite{ATLAS:2021kxv} and the LEP $Z$-pair~\cite{Electroweak:2003ram} and mono-photon~\cite{L3:2003yon} constraints to derive a lower bound on sneutrino mass. First, let us recast the LEP $Z$-pair data, letting the $Z$-pair decay into $\tau^+\tau^-\mu^+\mu^-$ final state, which is the same as our signal from sneutrino pair, and allows us to derive a lower bound on the sneutrino mass, since the measured cross-section at LEP was found to be close to the SM expectation. However, we find that the resulting lower bound on the sneutrino mass is about 100 GeV, which is entirely within the current 13~TeV LHC exclusion (inside the red-shaded region in Fig.~\ref{fig:results}). This seems reasonable because the center-of-mass energy of LEP is only 209 GeV and sneutrino pair-production via the $Z$-boson (similar to Fig.~\ref{fig:signal}, but replacing the $q\bar{q}$ with $e^+e^-$) is kinematically suppressed for sneutrino masses beyond $\sim 100$ GeV. Similarly, we find that the recast mono-photon bound from LEP for the channel $e^+e^-\to \widetilde{\nu}_\tau \widetilde{\nu}_\tau^* \to \widetilde{\chi}_1^0 \overline{\widetilde{\chi}}_1^0 \nu \bar{\nu}$ with an initial-state-radiation of photon is always weaker than the model-independent limit on sneutrino mass of 41 GeV because the experimental uncertainty of the measured cross section~\cite{L3:2003yon} is relatively large. Similarly, the mono-jet bound from LHC for the channel $pp\to \widetilde{\nu}_\tau \widetilde{\nu}_\tau^* \to \widetilde{\chi}_1^0 \overline{\widetilde{\chi}}_1^0 \nu \bar{\nu}$ with an initial-state-radiation of gluon is also weaker than the model-independent LEP limit used here due to small signal cross-section (in the absence of any $\lambda'$ couplings). For these reasons, the collider 
constraints we derived 
in Fig.~\ref{fig:results} are the strongest so far.

We also note that Ref.~\cite{ATLAS:2021yyr} considered the cascade decay of sleptons via the neutralino and derived stringent bounds on the sneutrino mass up to 850 GeV, depending on the neutralino mass. Naively, it looks like our scenario (b) is within their exclusion curve.  However, we would like to stress that in the ATLAS analysis~\cite{ATLAS:2021yyr}, a mass-degeneracy of charged sleptons and sneutrinos of all three generations is assumed. This assumption introduces many more production and decay channels and makes the cross-section much larger. In our scenario, only the third-generation sneutrino is light (sub-TeV scale), while the others are decoupled. Therefore, the exclusion limits of Ref.~~\cite{ATLAS:2021yyr} cannot be directly compared to our results. Moreover, their results do not cover our scenarios (a) and (c).

\section{Neutralino Decay}
In the above discussion, the lightest supersymmetric particle (LSP) is assumed to be either the lightest neutralino $\widetilde{\chi}_1^0$ or the tau sneutrino $\widetilde{\nu}_\tau$. For $m_{\widetilde{\chi}_1^0}>m_{\widetilde{\nu}_\tau}$, the neutralino undergoes prompt decay into $\nu_\tau\widetilde{\nu}_\tau$ via its gauge coupling. On the other hand, for $m_{\widetilde{\chi}_1^0}<m_{\widetilde{\nu}_\tau}$, it  undergoes a three-body decay into $\mu^- \tau^+ \nu_\tau$ via an off-shell $\widetilde{\nu}_\tau$, with the corresponding decay width given by 
\begin{align}
    \Gamma(\widetilde{\chi}_1^0\to \mu^- \tau^+ \nu_\tau) & \simeq \frac{g^2|\lambda_{233}|^2}{512\pi^3}\frac{m_{\widetilde{\chi}_1^0}^5}{m_{\widetilde{\nu}_\tau}^4} .
\end{align}
This leads to a typical decay length of 
\begin{align}
    \tau(\widetilde{\chi}_1^0\to \mu^- \tau^+ \nu_\tau)\simeq \frac{20~{\rm cm}}{|\lambda_{233}|^2}\left( \frac{1~{\rm GeV}}{m_{\widetilde{\chi}_1^0}}\right)^5\left(\frac{m_{\widetilde{\nu}_\tau}}{400~{\rm GeV}}\right)^4,
\end{align}
which means that the decay can be either prompt or displaced, depending on the mass and coupling values. 

For $m_{\widetilde{\chi}_1^0} < m_\tau + m_\mu$, we have the loop-induced decay $\widetilde{\chi}_1^0\to \gamma+\nu (\bar\nu)$, with the decay width given by~\cite{Hall:1983id, Haber:1988px, Dreiner:2022swd}
\begin{align}
    \Gamma(\widetilde{\chi}_1^0\to \gamma\nu) \simeq \frac{|\lambda_{233}|^2\alpha^2m^3_{\widetilde{\chi}_1^0}}{512\pi^3\cos^2\theta_w}\left[ \frac{3m_\tau}{m_{\widetilde{\tau}}^2}\left(1+\log\frac{m_\tau^2}{m_{\widetilde{\tau}}^2}\right)\right]^2 \,
\end{align}
where $\alpha$ is the fine-structure constant and $\theta_w$ is the weak mixing angle. This decay mode is suppressed by the heavy stau mass (which is required to be heavier than 5.8 TeV in our case), with the corresponding decay length given by 
\begin{align}
    \tau(\widetilde{\chi}_1^0\to \gamma\nu+\gamma\bar{\nu}) \simeq \frac{10^6~{\rm cm}}{|\lambda_{233}|^2}\left( \frac{1~{\rm GeV}}{m_{\widetilde{\chi}_1^0}}\right)^3\left(\frac{m_{\widetilde{\tau}}}{6~{\rm TeV}}\right)^4,
\end{align}
which necessarily makes it long-lived. 

If gravitino is the LSP (and a potential dark matter candidate), then there is another possible decay mode for the neutralino into gravitino and photon~\cite{Covi:2009bk}:
\begin{align}
    \Gamma(\widetilde{\chi}_1^0\to \gamma\widetilde{G}) \simeq \frac{\cos^2\theta_w}{48\pi M_{\rm Pl}^2}\frac{m^3_{\widetilde{\chi}_1^0}}{x^2_{3/2}}\left(1-x_{3/2}^2\right)^3\left(1+3x^2_{3/2}\right),
\end{align}
where $x_{3/2}\equiv m_{\widetilde{G}}/m_{\widetilde{\chi}_1^0}$. However, this decay mode is suppressed by the square of the Planck mass $M_{\rm Pl}$, and again, makes the neutralino very long-lived. 


%

\section{Conclusions} \label{sec:conclusions}
The RPV3 framework provides a compelling solution to the persistent hints of lepton flavor universality violation. In this paper, we have proposed a new RPV3 solution to the muon $(g-2)$-anomaly using the $\lambda_{233}$ coupling. This is consistent with the low-energy flavor constraints and existing collider bounds. 
The scenario is also compatible with the $R_{K^{(*)}}$ and $R_{D^{(*)}}$ anomalies whether or not they survive in the end.\footnote{The latest LHCb results~\cite{LHCb:2022qnv, LHCb:2022zom} seem to indicate that the $R_{K^{(*)}}$ anomaly does not exist anymore.}

For the scenario under consideration, we have constructed new LHC bounds, following an existing ATLAS multi-lepton analysis with the Run-2 data. We have also shown how the bounds would improve with a dedicated selection of only two muons as the light leptons. The HL-LHC prospects were also discussed in this context.

We found that under the current LHC data, the muon $(g-2)$-favored region survives only for $m_{\widetilde{\nu}_\tau} \gtrsim 400$~GeV and $\lambda_{233} \gtrsim 2$. Unlike our previous results for the $\lambda_{232}\neq 0$~\cite{Dev:2021ipu}, where $m_{\widetilde{\nu}_\tau}$ was required to be larger than $ \gtrsim 650$~GeV, our new scenario allows lighter
sneutrinos. This is because the $\widetilde{\nu}_\tau$ decays into a $\mu\tau$-pair for $\lambda_{233}\neq 0$ rather than a $\mu\mu$ pair for $\lambda_{232}\neq 0$, and taus are more difficult than muons to identify experimentally.

The collider signal of $\mu^+\mu^-\tau^+\tau^-$ that we analyzed here is a generic prediction of {\it any} BSM scenario trying to explain the muon $(g-2)$ via a tau-loop, either with or without chirality enhancement. Therefore, the analysis presented here can be extended to all such models, although the specific details, such as the signal cross section or the $(g-2)$-preferred range of model parameters, might be somewhat different.  

We expect new results forthcoming from the Fermilab Muon $(g-2)$ experiment, as a lot more data has been accumulated since the first results were announced in 2021. Another muon $(g-2)$ experiment with similar sensitivity but using a different technique is currently under construction at J-PARC~\cite{Abe:2019thb}. On the theory front, more refined SM calculations for $a_\mu$ are currently underway~\cite{Colangelo:2022jxc}. An independent measurement of the leading order hadronic contribution to $a_\mu$ has also been proposed from the MuonE experiment at CERN~\cite{Abbiendi:2016xup}, which is immune to any possible BSM contamination~\cite{Dev:2020drf,Masiero:2020vxk}. All in all, it is very likely that the fate of the muon $(g-2)$ anomaly will be sealed beyond a reasonable doubt in the not-so-distant future. Our proposed collider signal will {\it independently} test the BSM interpretation of the muon $(g-2)$ anomaly in any model with lepton flavor violating $\mu\tau$ couplings. This may also have implications for lepton flavor universality tests in the $B$-meson decays.    

\begin{figure*}[ht!]
    \centering
    \subfloat[]{\includegraphics[width=0.49\textwidth]{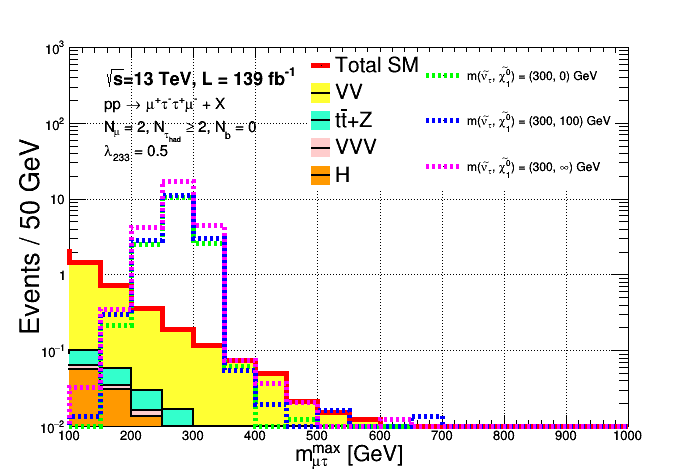}}
    \subfloat[]{\includegraphics[width=0.49\textwidth]{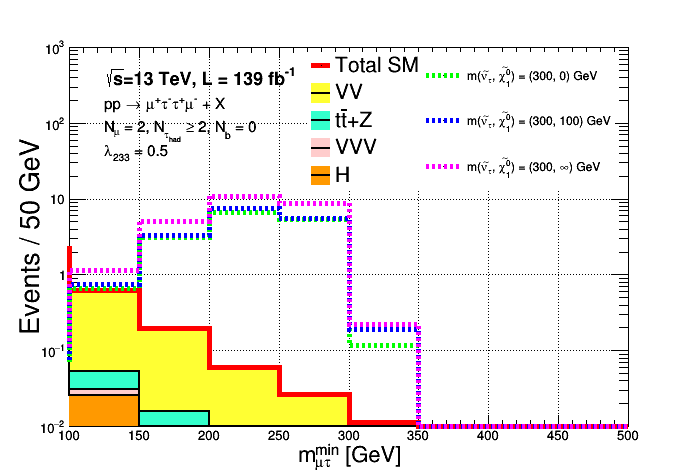}}\\
    \subfloat[]{\includegraphics[width=0.49\textwidth]{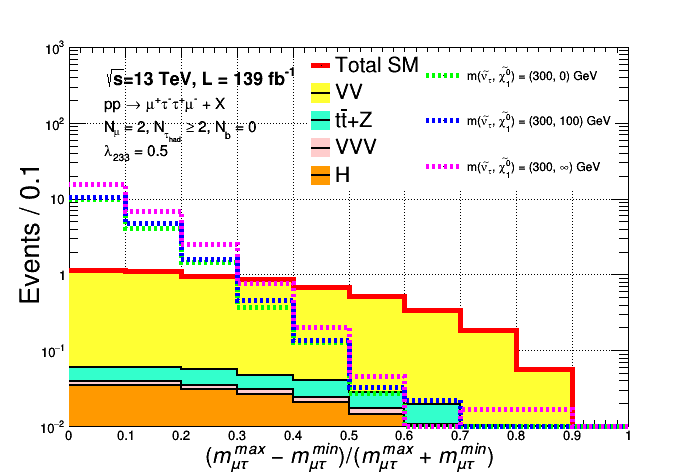}}
    \caption{Additional distributions that can be used in a dedicated analysis: (a) $m_\mathrm{\mu\tau}^\mathrm{max}$, (b) $m_\mathrm{\mu\tau}^\mathrm{min}$ and (c) $(m_\mathrm{\mu\tau}^\mathrm{max} - m_\mathrm{\mu\tau}^\mathrm{min}) / (m_\mathrm{\mu\tau}^\mathrm{max} + m_\mathrm{\mu\tau}^\mathrm{min})$. All of the selections of $\text{SR2}_{\text{bveto}}^{\text{loose}}$-$\mu\mu$ and $\text{SR2}_{\text{bveto}}^{\text{tight}}$-$\mu\mu$ are implemented, as described in Tab.~\ref{tab:selections}, beside the $m_\mathrm{eff}$ selection. Three signal points are presented by setting $m_{\widetilde{\nu}_\tau}=300$~GeV
    and $\lambda_{233}=0.5$, for three different choices of $m_{\widetilde{\chi}_1^0}$: with a very small value (0~GeV), 100~GeV, and a very large value ($\infty$).
    }
    \label{fig:distributions}
\end{figure*}
\acknowledgments
BD is grateful to Lawrence Hall for illuminating discussions during a seminar visit to UC Berkeley in March 2022 that inspired this analysis. YA thanks Rosa Simoniello for directing him to the most recent and relevant experimental results. AS is very grateful to Christoph Lehner for discussions pertaining to  muon $(g-2)$. 
The work of BD and FX is supported in part by the US Department of Energy under Grant No.~DE-SC0017987 and by the MCSS funds. BD is also supported in part by a URA VSP fellowship. The work of AS  was supported in part by the U.S. DOE contract \#DE-SC0012704.
\appendix
\section{Kinematic Distributions} \label{sec:kinematic}
Additional kinematic observables can be used in order to enhance the sensitivity for a given model.
For the RPV3 model considered in this work, we list a few of these and show the kinematic distributions of the signal (for three benchmark cases) and the SM backgrounds (shaded histograms) in Fig.~\ref{fig:distributions}. This is to emphasize the future potential of the $\mu^+\mu^-\tau^+\tau^-$ channel at the HL-LHC. The variables considered here are:
\begin{itemize}
    \item $m_\mathrm{\mu\tau}^\mathrm{max}$: the maximum value of the invariant mass of a pair of a muon and the visible part of a hadronically decaying tau-lepton, $\tau_\mathrm{had}^\mathrm{vis}$, with opposite charges.\footnote{The charge of the hadronically decaying tau-lepton can be identified by the sum of charges of its decay products.} This observable is expected to peak close to the mass of the sneutrino for the signal.
    \item $m_\mathrm{\mu\tau}^\mathrm{min}$: similar to the observable above, but with the minimum value instead.
    \item $(m_\mathrm{\mu\tau}^\mathrm{max} - m_\mathrm{\mu\tau}^\mathrm{min}) / (m_\mathrm{\mu\tau}^\mathrm{max} + m_\mathrm{\mu\tau}^\mathrm{min})$: since the signal production includes two resonances with similar masses, we expect that the difference between the invariant masses of their decay products will be similar, different than the SM backgrounds.

\end{itemize}

\bibliography{mybib2}

\begin{thebibliography}{89}%
\makeatletter
\providecommand \@ifxundefined [1]{%
 \@ifx{#1\undefined}
}%
\providecommand \@ifnum [1]{%
 \ifnum #1\expandafter \@firstoftwo
 \else \expandafter \@secondoftwo
 \fi
}%
\providecommand \@ifx [1]{%
 \ifx #1\expandafter \@firstoftwo
 \else \expandafter \@secondoftwo
 \fi
}%
\providecommand \natexlab [1]{#1}%
\providecommand \enquote  [1]{``#1''}%
\providecommand \bibnamefont  [1]{#1}%
\providecommand \bibfnamefont [1]{#1}%
\providecommand \citenamefont [1]{#1}%
\providecommand \href@noop [0]{\@secondoftwo}%
\providecommand \href [0]{\begingroup \@sanitize@url \@href}%
\providecommand \@href[1]{\@@startlink{#1}\@@href}%
\providecommand \@@href[1]{\endgroup#1\@@endlink}%
\providecommand \@sanitize@url [0]{\catcode `\\12\catcode `\$12\catcode
  `\&12\catcode `\#12\catcode `\^12\catcode `\_12\catcode `\%12\relax}%
\providecommand \@@startlink[1]{}%
\providecommand \@@endlink[0]{}%
\providecommand \url  [0]{\begingroup\@sanitize@url \@url }%
\providecommand \@url [1]{\endgroup\@href {#1}{\urlprefix }}%
\providecommand \urlprefix  [0]{URL }%
\providecommand \Eprint [0]{\href }%
\providecommand \doibase [0]{https://doi.org/}%
\providecommand \selectlanguage [0]{\@gobble}%
\providecommand \bibinfo  [0]{\@secondoftwo}%
\providecommand \bibfield  [0]{\@secondoftwo}%
\providecommand \translation [1]{[#1]}%
\providecommand \BibitemOpen [0]{}%
\providecommand \bibitemStop [0]{}%
\providecommand \bibitemNoStop [0]{.\EOS\space}%
\providecommand \EOS [0]{\spacefactor3000\relax}%
\providecommand \BibitemShut  [1]{\csname bibitem#1\endcsname}%
\let\auto@bib@innerbib\@empty
\bibitem [{\citenamefont {Workman}\ \emph {et~al.}(2022)\citenamefont {Workman}
  \emph {et~al.}}]{Workman:2022ynf}%
  \BibitemOpen
  \bibfield  {author} {\bibinfo {author} {\bibfnamefont {R.~L.}\ \bibnamefont
  {Workman}} \emph {et~al.} (\bibinfo {collaboration} {Particle Data Group}),\
  }\bibfield  {title} {\bibinfo {title} {{Review of Particle Physics}},\ }\href
  {https://doi.org/10.1093/ptep/ptac097} {\bibfield  {journal} {\bibinfo
  {journal} {PTEP}\ }\textbf {\bibinfo {volume} {2022}},\ \bibinfo {pages}
  {083C01} (\bibinfo {year} {2022})}\BibitemShut {NoStop}%
\bibitem [{\citenamefont {Bennett}\ \emph {et~al.}(2006)\citenamefont {Bennett}
  \emph {et~al.}}]{Muong-2:2006rrc}%
  \BibitemOpen
  \bibfield  {author} {\bibinfo {author} {\bibfnamefont {G.~W.}\ \bibnamefont
  {Bennett}} \emph {et~al.} (\bibinfo {collaboration} {Muon g-2}),\ }\bibfield
  {title} {\bibinfo {title} {{Final Report of the Muon E821 Anomalous Magnetic
  Moment Measurement at BNL}},\ }\href
  {https://doi.org/10.1103/PhysRevD.73.072003} {\bibfield  {journal} {\bibinfo
  {journal} {Phys. Rev. D}\ }\textbf {\bibinfo {volume} {73}},\ \bibinfo
  {pages} {072003} (\bibinfo {year} {2006})},\ \Eprint
  {https://arxiv.org/abs/hep-ex/0602035} {arXiv:hep-ex/0602035} \BibitemShut
  {NoStop}%
\bibitem [{\citenamefont {Abi}\ \emph {et~al.}(2021)\citenamefont {Abi} \emph
  {et~al.}}]{Muong-2:2021ojo}%
  \BibitemOpen
  \bibfield  {author} {\bibinfo {author} {\bibfnamefont {B.}~\bibnamefont
  {Abi}} \emph {et~al.} (\bibinfo {collaboration} {Muon g-2}),\ }\bibfield
  {title} {\bibinfo {title} {{Measurement of the Positive Muon Anomalous
  Magnetic Moment to 0.46 ppm}},\ }\href
  {https://doi.org/10.1103/PhysRevLett.126.141801} {\bibfield  {journal}
  {\bibinfo  {journal} {Phys. Rev. Lett.}\ }\textbf {\bibinfo {volume} {126}},\
  \bibinfo {pages} {141801} (\bibinfo {year} {2021})},\ \Eprint
  {https://arxiv.org/abs/2104.03281} {arXiv:2104.03281 [hep-ex]} \BibitemShut
  {NoStop}%
\bibitem [{\citenamefont {Aoyama}\ \emph {et~al.}(2020)\citenamefont {Aoyama}
  \emph {et~al.}}]{Aoyama:2020ynm}%
  \BibitemOpen
  \bibfield  {author} {\bibinfo {author} {\bibfnamefont {T.}~\bibnamefont
  {Aoyama}} \emph {et~al.},\ }\bibfield  {title} {\bibinfo {title} {{The
  anomalous magnetic moment of the muon in the Standard Model}},\ }\href
  {https://doi.org/10.1016/j.physrep.2020.07.006} {\bibfield  {journal}
  {\bibinfo  {journal} {Phys. Rept.}\ }\textbf {\bibinfo {volume} {887}},\
  \bibinfo {pages} {1} (\bibinfo {year} {2020})},\ \Eprint
  {https://arxiv.org/abs/2006.04822} {arXiv:2006.04822 [hep-ph]} \BibitemShut
  {NoStop}%
\bibitem [{\citenamefont {Borsanyi}\ \emph {et~al.}(2021)\citenamefont
  {Borsanyi} \emph {et~al.}}]{Borsanyi:2020mff}%
  \BibitemOpen
  \bibfield  {author} {\bibinfo {author} {\bibfnamefont {S.}~\bibnamefont
  {Borsanyi}} \emph {et~al.},\ }\bibfield  {title} {\bibinfo {title} {{Leading
  hadronic contribution to the muon magnetic moment from lattice QCD}},\ }\href
  {https://doi.org/10.1038/s41586-021-03418-1} {\bibfield  {journal} {\bibinfo
  {journal} {Nature}\ }\textbf {\bibinfo {volume} {593}},\ \bibinfo {pages}
  {51} (\bibinfo {year} {2021})},\ \Eprint {https://arxiv.org/abs/2002.12347}
  {arXiv:2002.12347 [hep-lat]} \BibitemShut {NoStop}%
\bibitem [{\citenamefont {C\`e}\ \emph {et~al.}(2022)\citenamefont {C\`e} \emph
  {et~al.}}]{Ce:2022kxy}%
  \BibitemOpen
  \bibfield  {author} {\bibinfo {author} {\bibfnamefont {M.}~\bibnamefont
  {C\`e}} \emph {et~al.},\ }\bibfield  {title} {\bibinfo {title} {{Window
  observable for the hadronic vacuum polarization contribution to the muon
  $g-2$ from lattice QCD}},\ }\href@noop {} {\  (\bibinfo {year} {2022})},\
  \Eprint {https://arxiv.org/abs/2206.06582} {arXiv:2206.06582 [hep-lat]}
  \BibitemShut {NoStop}%
\bibitem [{\citenamefont {Alexandrou}\ \emph {et~al.}(2022)\citenamefont
  {Alexandrou} \emph {et~al.}}]{Alexandrou:2022amy}%
  \BibitemOpen
  \bibfield  {author} {\bibinfo {author} {\bibfnamefont {C.}~\bibnamefont
  {Alexandrou}} \emph {et~al.},\ }\bibfield  {title} {\bibinfo {title}
  {{Lattice calculation of the short and intermediate time-distance hadronic
  vacuum polarization contributions to the muon magnetic moment using
  twisted-mass fermions}},\ }\href@noop {} {\  (\bibinfo {year} {2022})},\
  \Eprint {https://arxiv.org/abs/2206.15084} {arXiv:2206.15084 [hep-lat]}
  \BibitemShut {NoStop}%
\bibitem [{\citenamefont {Colangelo}\ \emph
  {et~al.}(2022{\natexlab{a}})\citenamefont {Colangelo}, \citenamefont
  {El-Khadra}, \citenamefont {Hoferichter}, \citenamefont {Keshavarzi},
  \citenamefont {Lehner}, \citenamefont {Stoffer},\ and\ \citenamefont
  {Teubner}}]{Colangelo:2022vok}%
  \BibitemOpen
  \bibfield  {author} {\bibinfo {author} {\bibfnamefont {G.}~\bibnamefont
  {Colangelo}}, \bibinfo {author} {\bibfnamefont {A.~X.}\ \bibnamefont
  {El-Khadra}}, \bibinfo {author} {\bibfnamefont {M.}~\bibnamefont
  {Hoferichter}}, \bibinfo {author} {\bibfnamefont {A.}~\bibnamefont
  {Keshavarzi}}, \bibinfo {author} {\bibfnamefont {C.}~\bibnamefont {Lehner}},
  \bibinfo {author} {\bibfnamefont {P.}~\bibnamefont {Stoffer}},\ and\ \bibinfo
  {author} {\bibfnamefont {T.}~\bibnamefont {Teubner}},\ }\bibfield  {title}
  {\bibinfo {title} {{Data-driven evaluations of Euclidean windows to
  scrutinize hadronic vacuum polarization}},\ }\href
  {https://doi.org/10.1016/j.physletb.2022.137313} {\bibfield  {journal}
  {\bibinfo  {journal} {Phys. Lett. B}\ }\textbf {\bibinfo {volume} {833}},\
  \bibinfo {pages} {137313} (\bibinfo {year} {2022}{\natexlab{a}})},\ \Eprint
  {https://arxiv.org/abs/2205.12963} {arXiv:2205.12963 [hep-ph]} \BibitemShut
  {NoStop}%
\bibitem [{\citenamefont {Lehner}(2022)}]{talkLehner2022}%
  \BibitemOpen
  \bibfield  {author} {\bibinfo {author} {\bibfnamefont {C.}~\bibnamefont
  {Lehner}} (\bibinfo {collaboration} {RBC and UKQCD}),\ }\href@noop {}
  {\bibinfo {title} {The hadronic vacuum polarization}},\ \bibinfo
  {howpublished}
  {\url{https://indico.ph.ed.ac.uk/event/112/contributions/1660/attachments/1000/1391/talk-nobackup.pdf}}
  (\bibinfo {year} {2022}),\ \bibinfo {note} {fifth Plenary Workshop of the
  Muon g-2 Theory Initiative, Edinburgh, UK}\BibitemShut {NoStop}%
\bibitem [{\citenamefont {Gottlieb}(2022)}]{talkGottlieb2022}%
  \BibitemOpen
  \bibfield  {author} {\bibinfo {author} {\bibfnamefont {S.}~\bibnamefont
  {Gottlieb}} (\bibinfo {collaboration} {Fermilab Lattice, HPQCD and MILC}),\
  }\href@noop {} {\bibinfo {title} {Hadronic vacuum polarization: An unblinded
  window on the g-2 mystery}},\ \bibinfo {howpublished}
  {\url{https://www.benasque.org/2022lattice_workshop/talks_contr/158_Gottlieb_gm2_LatticeNET.pdf}}
  (\bibinfo {year} {2022}),\ \bibinfo {note} {first LatticeNET Workshop on
  challenges in Lattice field theory, Benasque, Spain}\BibitemShut {NoStop}%
\bibitem [{\citenamefont {Colangelo}(2022)}]{talkColangelo2022}%
  \BibitemOpen
  \bibfield  {author} {\bibinfo {author} {\bibfnamefont {G.}~\bibnamefont
  {Colangelo}},\ }\href@noop {} {\bibinfo {title} {Dispersive calculation of
  hadronic contributions to $(g-2)_\mu$}},\ \bibinfo {howpublished}
  {\url{https://www.benasque.org/2022lattice_workshop/talks_contr/153_g-2_Benasque-2022.pdf}}
  (\bibinfo {year} {2022}),\ \bibinfo {note} {first LatticeNET Workshop on
  challenges in Lattice field theory, Benasque, Spain}\BibitemShut {NoStop}%
\bibitem [{\citenamefont {Crivellin}\ \emph {et~al.}(2020)\citenamefont
  {Crivellin}, \citenamefont {Hoferichter}, \citenamefont {Manzari},\ and\
  \citenamefont {Montull}}]{Crivellin:2020zul}%
  \BibitemOpen
  \bibfield  {author} {\bibinfo {author} {\bibfnamefont {A.}~\bibnamefont
  {Crivellin}}, \bibinfo {author} {\bibfnamefont {M.}~\bibnamefont
  {Hoferichter}}, \bibinfo {author} {\bibfnamefont {C.~A.}\ \bibnamefont
  {Manzari}},\ and\ \bibinfo {author} {\bibfnamefont {M.}~\bibnamefont
  {Montull}},\ }\bibfield  {title} {\bibinfo {title} {{Hadronic Vacuum
  Polarization: $(g-2)_\mu$ versus Global Electroweak Fits}},\ }\href
  {https://doi.org/10.1103/PhysRevLett.125.091801} {\bibfield  {journal}
  {\bibinfo  {journal} {Phys. Rev. Lett.}\ }\textbf {\bibinfo {volume} {125}},\
  \bibinfo {pages} {091801} (\bibinfo {year} {2020})},\ \Eprint
  {https://arxiv.org/abs/2003.04886} {arXiv:2003.04886 [hep-ph]} \BibitemShut
  {NoStop}%
\bibitem [{\citenamefont {Keshavarzi}\ \emph {et~al.}(2020)\citenamefont
  {Keshavarzi}, \citenamefont {Marciano}, \citenamefont {Passera},\ and\
  \citenamefont {Sirlin}}]{Keshavarzi:2020bfy}%
  \BibitemOpen
  \bibfield  {author} {\bibinfo {author} {\bibfnamefont {A.}~\bibnamefont
  {Keshavarzi}}, \bibinfo {author} {\bibfnamefont {W.~J.}\ \bibnamefont
  {Marciano}}, \bibinfo {author} {\bibfnamefont {M.}~\bibnamefont {Passera}},\
  and\ \bibinfo {author} {\bibfnamefont {A.}~\bibnamefont {Sirlin}},\
  }\bibfield  {title} {\bibinfo {title} {{Muon $g-2$ and $\Delta \alpha$
  connection}},\ }\href {https://doi.org/10.1103/PhysRevD.102.033002}
  {\bibfield  {journal} {\bibinfo  {journal} {Phys. Rev. D}\ }\textbf {\bibinfo
  {volume} {102}},\ \bibinfo {pages} {033002} (\bibinfo {year} {2020})},\
  \Eprint {https://arxiv.org/abs/2006.12666} {arXiv:2006.12666 [hep-ph]}
  \BibitemShut {NoStop}%
\bibitem [{\citenamefont {Colangelo}\ \emph {et~al.}(2021)\citenamefont
  {Colangelo}, \citenamefont {Hoferichter},\ and\ \citenamefont
  {Stoffer}}]{Colangelo:2020lcg}%
  \BibitemOpen
  \bibfield  {author} {\bibinfo {author} {\bibfnamefont {G.}~\bibnamefont
  {Colangelo}}, \bibinfo {author} {\bibfnamefont {M.}~\bibnamefont
  {Hoferichter}},\ and\ \bibinfo {author} {\bibfnamefont {P.}~\bibnamefont
  {Stoffer}},\ }\bibfield  {title} {\bibinfo {title} {{Constraints on the
  two-pion contribution to hadronic vacuum polarization}},\ }\href
  {https://doi.org/10.1016/j.physletb.2021.136073} {\bibfield  {journal}
  {\bibinfo  {journal} {Phys. Lett. B}\ }\textbf {\bibinfo {volume} {814}},\
  \bibinfo {pages} {136073} (\bibinfo {year} {2021})},\ \Eprint
  {https://arxiv.org/abs/2010.07943} {arXiv:2010.07943 [hep-ph]} \BibitemShut
  {NoStop}%
\bibitem [{\citenamefont {Lindner}\ \emph {et~al.}(2018)\citenamefont
  {Lindner}, \citenamefont {Platscher},\ and\ \citenamefont
  {Queiroz}}]{Lindner:2016bgg}%
  \BibitemOpen
  \bibfield  {author} {\bibinfo {author} {\bibfnamefont {M.}~\bibnamefont
  {Lindner}}, \bibinfo {author} {\bibfnamefont {M.}~\bibnamefont {Platscher}},\
  and\ \bibinfo {author} {\bibfnamefont {F.~S.}\ \bibnamefont {Queiroz}},\
  }\bibfield  {title} {\bibinfo {title} {{A Call for New Physics : The Muon
  Anomalous Magnetic Moment and Lepton Flavor Violation}},\ }\href
  {https://doi.org/10.1016/j.physrep.2017.12.001} {\bibfield  {journal}
  {\bibinfo  {journal} {Phys. Rept.}\ }\textbf {\bibinfo {volume} {731}},\
  \bibinfo {pages} {1} (\bibinfo {year} {2018})},\ \Eprint
  {https://arxiv.org/abs/1610.06587} {arXiv:1610.06587 [hep-ph]} \BibitemShut
  {NoStop}%
\bibitem [{\citenamefont {Athron}\ \emph {et~al.}(2021)\citenamefont {Athron},
  \citenamefont {Bal\'azs}, \citenamefont {Jacob}, \citenamefont {Kotlarski},
  \citenamefont {St\"ockinger},\ and\ \citenamefont
  {St\"ockinger-Kim}}]{Athron:2021iuf}%
  \BibitemOpen
  \bibfield  {author} {\bibinfo {author} {\bibfnamefont {P.}~\bibnamefont
  {Athron}}, \bibinfo {author} {\bibfnamefont {C.}~\bibnamefont {Bal\'azs}},
  \bibinfo {author} {\bibfnamefont {D.~H.~J.}\ \bibnamefont {Jacob}}, \bibinfo
  {author} {\bibfnamefont {W.}~\bibnamefont {Kotlarski}}, \bibinfo {author}
  {\bibfnamefont {D.}~\bibnamefont {St\"ockinger}},\ and\ \bibinfo {author}
  {\bibfnamefont {H.}~\bibnamefont {St\"ockinger-Kim}},\ }\bibfield  {title}
  {\bibinfo {title} {{New physics explanations of a$_{\mu}$ in light of the
  FNAL muon $g-2$ measurement}},\ }\href
  {https://doi.org/10.1007/JHEP09(2021)080} {\bibfield  {journal} {\bibinfo
  {journal} {JHEP}\ }\textbf {\bibinfo {volume} {09}},\ \bibinfo {pages}
  {080}},\ \Eprint {https://arxiv.org/abs/2104.03691} {arXiv:2104.03691
  [hep-ph]} \BibitemShut {NoStop}%
\bibitem [{\citenamefont {Afik}\ \emph {et~al.}(2022)\citenamefont {Afik},
  \citenamefont {Bar-Shalom}, \citenamefont {Pal}, \citenamefont {Soni},\ and\
  \citenamefont {Wudka}}]{Afik:2021xmi}%
  \BibitemOpen
  \bibfield  {author} {\bibinfo {author} {\bibfnamefont {Y.}~\bibnamefont
  {Afik}}, \bibinfo {author} {\bibfnamefont {S.}~\bibnamefont {Bar-Shalom}},
  \bibinfo {author} {\bibfnamefont {K.}~\bibnamefont {Pal}}, \bibinfo {author}
  {\bibfnamefont {A.}~\bibnamefont {Soni}},\ and\ \bibinfo {author}
  {\bibfnamefont {J.}~\bibnamefont {Wudka}},\ }\bibfield  {title} {\bibinfo
  {title} {{Multi-lepton probes of new physics and lepton-universality in
  top-quark interactions}},\ }\href
  {https://doi.org/10.1016/j.nuclphysb.2022.115849} {\bibfield  {journal}
  {\bibinfo  {journal} {Nucl. Phys. B}\ }\textbf {\bibinfo {volume} {980}},\
  \bibinfo {pages} {115849} (\bibinfo {year} {2022})},\ \Eprint
  {https://arxiv.org/abs/2111.13711} {arXiv:2111.13711 [hep-ph]} \BibitemShut
  {NoStop}%
\bibitem [{\citenamefont {Jackiw}\ and\ \citenamefont
  {Weinberg}(1972)}]{Jackiw:1972jz}%
  \BibitemOpen
  \bibfield  {author} {\bibinfo {author} {\bibfnamefont {R.}~\bibnamefont
  {Jackiw}}\ and\ \bibinfo {author} {\bibfnamefont {S.}~\bibnamefont
  {Weinberg}},\ }\bibfield  {title} {\bibinfo {title} {{Weak interaction
  corrections to the muon magnetic moment and to muonic atom energy levels}},\
  }\href {https://doi.org/10.1103/PhysRevD.5.2396} {\bibfield  {journal}
  {\bibinfo  {journal} {Phys. Rev. D}\ }\textbf {\bibinfo {volume} {5}},\
  \bibinfo {pages} {2396} (\bibinfo {year} {1972})}\BibitemShut {NoStop}%
\bibitem [{\citenamefont {Czarnecki}\ and\ \citenamefont
  {Marciano}(2001)}]{Czarnecki:2001pv}%
  \BibitemOpen
  \bibfield  {author} {\bibinfo {author} {\bibfnamefont {A.}~\bibnamefont
  {Czarnecki}}\ and\ \bibinfo {author} {\bibfnamefont {W.~J.}\ \bibnamefont
  {Marciano}},\ }\bibfield  {title} {\bibinfo {title} {{The Muon anomalous
  magnetic moment: A Harbinger for 'new physics'}},\ }\href
  {https://doi.org/10.1103/PhysRevD.64.013014} {\bibfield  {journal} {\bibinfo
  {journal} {Phys. Rev. D}\ }\textbf {\bibinfo {volume} {64}},\ \bibinfo
  {pages} {013014} (\bibinfo {year} {2001})},\ \Eprint
  {https://arxiv.org/abs/hep-ph/0102122} {arXiv:hep-ph/0102122} \BibitemShut
  {NoStop}%
\bibitem [{\citenamefont {Crivellin}\ and\ \citenamefont
  {Hoferichter}(2022)}]{Crivellin:2022wzw}%
  \BibitemOpen
  \bibfield  {author} {\bibinfo {author} {\bibfnamefont {A.}~\bibnamefont
  {Crivellin}}\ and\ \bibinfo {author} {\bibfnamefont {M.}~\bibnamefont
  {Hoferichter}},\ }\bibfield  {title} {\bibinfo {title} {{The Anomalous
  Magnetic Moment of the Muon: Beyond the Standard Model via Chiral
  Enhancement}}\ }(\bibinfo {year} {2022})\ \Eprint
  {https://arxiv.org/abs/2207.01912} {arXiv:2207.01912 [hep-ph]} \BibitemShut
  {NoStop}%
\bibitem [{\citenamefont {St\"ockinger}\ and\ \citenamefont
  {St\"ockinger-Kim}(2022)}]{Stockinger:2022ata}%
  \BibitemOpen
  \bibfield  {author} {\bibinfo {author} {\bibfnamefont {D.}~\bibnamefont
  {St\"ockinger}}\ and\ \bibinfo {author} {\bibfnamefont {H.}~\bibnamefont
  {St\"ockinger-Kim}},\ }\bibfield  {title} {\bibinfo {title} {{On the role of
  chirality flips for the muon magnetic moment and its relation to the muon
  mass}},\ }\href {https://doi.org/10.3389/fphy.2022.944614} {\bibfield
  {journal} {\bibinfo  {journal} {Front. in Phys.}\ }\textbf {\bibinfo {volume}
  {10}},\ \bibinfo {pages} {944614} (\bibinfo {year} {2022})}\BibitemShut
  {NoStop}%
\bibitem [{\citenamefont {Freitas}\ \emph {et~al.}(2014)\citenamefont
  {Freitas}, \citenamefont {Lykken}, \citenamefont {Kell},\ and\ \citenamefont
  {Westhoff}}]{Freitas:2014pua}%
  \BibitemOpen
  \bibfield  {author} {\bibinfo {author} {\bibfnamefont {A.}~\bibnamefont
  {Freitas}}, \bibinfo {author} {\bibfnamefont {J.}~\bibnamefont {Lykken}},
  \bibinfo {author} {\bibfnamefont {S.}~\bibnamefont {Kell}},\ and\ \bibinfo
  {author} {\bibfnamefont {S.}~\bibnamefont {Westhoff}},\ }\bibfield  {title}
  {\bibinfo {title} {{Testing the Muon g-2 Anomaly at the LHC}},\ }\href
  {https://doi.org/10.1007/JHEP09(2014)155} {\bibfield  {journal} {\bibinfo
  {journal} {JHEP}\ }\textbf {\bibinfo {volume} {05}},\ \bibinfo {pages}
  {145}},\ \bibinfo {note} {[Erratum: JHEP 09, 155 (2014)]},\ \Eprint
  {https://arxiv.org/abs/1402.7065} {arXiv:1402.7065 [hep-ph]} \BibitemShut
  {NoStop}%
\bibitem [{\citenamefont {Sabatta}\ \emph {et~al.}(2020)\citenamefont
  {Sabatta}, \citenamefont {Cornell}, \citenamefont {Goyal}, \citenamefont
  {Kumar}, \citenamefont {Mellado},\ and\ \citenamefont
  {Ruan}}]{Sabatta:2019nfg}%
  \BibitemOpen
  \bibfield  {author} {\bibinfo {author} {\bibfnamefont {D.}~\bibnamefont
  {Sabatta}}, \bibinfo {author} {\bibfnamefont {A.~S.}\ \bibnamefont
  {Cornell}}, \bibinfo {author} {\bibfnamefont {A.}~\bibnamefont {Goyal}},
  \bibinfo {author} {\bibfnamefont {M.}~\bibnamefont {Kumar}}, \bibinfo
  {author} {\bibfnamefont {B.}~\bibnamefont {Mellado}},\ and\ \bibinfo {author}
  {\bibfnamefont {X.}~\bibnamefont {Ruan}},\ }\bibfield  {title} {\bibinfo
  {title} {{Connecting muon anomalous magnetic moment and multi-lepton
  anomalies at LHC}},\ }\href {https://doi.org/10.1088/1674-1137/44/6/063103}
  {\bibfield  {journal} {\bibinfo  {journal} {Chin. Phys. C}\ }\textbf
  {\bibinfo {volume} {44}},\ \bibinfo {pages} {063103} (\bibinfo {year}
  {2020})},\ \Eprint {https://arxiv.org/abs/1909.03969} {arXiv:1909.03969
  [hep-ph]} \BibitemShut {NoStop}%
\bibitem [{\citenamefont {Capdevilla}\ \emph {et~al.}(2022)\citenamefont
  {Capdevilla}, \citenamefont {Curtin}, \citenamefont {Kahn},\ and\
  \citenamefont {Krnjaic}}]{Capdevilla:2021rwo}%
  \BibitemOpen
  \bibfield  {author} {\bibinfo {author} {\bibfnamefont {R.}~\bibnamefont
  {Capdevilla}}, \bibinfo {author} {\bibfnamefont {D.}~\bibnamefont {Curtin}},
  \bibinfo {author} {\bibfnamefont {Y.}~\bibnamefont {Kahn}},\ and\ \bibinfo
  {author} {\bibfnamefont {G.}~\bibnamefont {Krnjaic}},\ }\bibfield  {title}
  {\bibinfo {title} {{No-lose theorem for discovering the new physics of
  (g-2)\ensuremath{\mu} at muon colliders}},\ }\href
  {https://doi.org/10.1103/PhysRevD.105.015028} {\bibfield  {journal} {\bibinfo
   {journal} {Phys. Rev. D}\ }\textbf {\bibinfo {volume} {105}},\ \bibinfo
  {pages} {015028} (\bibinfo {year} {2022})},\ \Eprint
  {https://arxiv.org/abs/2101.10334} {arXiv:2101.10334 [hep-ph]} \BibitemShut
  {NoStop}%
\bibitem [{\citenamefont {Arkani-Hamed}\ and\ \citenamefont
  {Harigaya}(2021)}]{Arkani-Hamed:2021xlp}%
  \BibitemOpen
  \bibfield  {author} {\bibinfo {author} {\bibfnamefont {N.}~\bibnamefont
  {Arkani-Hamed}}\ and\ \bibinfo {author} {\bibfnamefont {K.}~\bibnamefont
  {Harigaya}},\ }\bibfield  {title} {\bibinfo {title} {{Naturalness and the
  muon magnetic moment}},\ }\href {https://doi.org/10.1007/JHEP09(2021)025}
  {\bibfield  {journal} {\bibinfo  {journal} {JHEP}\ }\textbf {\bibinfo
  {volume} {09}},\ \bibinfo {pages} {025}},\ \Eprint
  {https://arxiv.org/abs/2106.01373} {arXiv:2106.01373 [hep-ph]} \BibitemShut
  {NoStop}%
\bibitem [{\citenamefont {Babu}\ \emph {et~al.}(2020)\citenamefont {Babu},
  \citenamefont {Jana},\ and\ \citenamefont {Lindner}}]{Babu:2020ivd}%
  \BibitemOpen
  \bibfield  {author} {\bibinfo {author} {\bibfnamefont {K.~S.}\ \bibnamefont
  {Babu}}, \bibinfo {author} {\bibfnamefont {S.}~\bibnamefont {Jana}},\ and\
  \bibinfo {author} {\bibfnamefont {M.}~\bibnamefont {Lindner}},\ }\bibfield
  {title} {\bibinfo {title} {{Large Neutrino Magnetic Moments in the Light of
  Recent Experiments}},\ }\href {https://doi.org/10.1007/JHEP10(2020)040}
  {\bibfield  {journal} {\bibinfo  {journal} {JHEP}\ }\textbf {\bibinfo
  {volume} {10}},\ \bibinfo {pages} {040}},\ \Eprint
  {https://arxiv.org/abs/2007.04291} {arXiv:2007.04291 [hep-ph]} \BibitemShut
  {NoStop}%
\bibitem [{\citenamefont {Barbier}\ \emph {et~al.}(2005)\citenamefont {Barbier}
  \emph {et~al.}}]{Barbier:2004ez}%
  \BibitemOpen
  \bibfield  {author} {\bibinfo {author} {\bibfnamefont {R.}~\bibnamefont
  {Barbier}} \emph {et~al.},\ }\bibfield  {title} {\bibinfo {title} {{R-parity
  violating supersymmetry}},\ }\href
  {https://doi.org/10.1016/j.physrep.2005.08.006} {\bibfield  {journal}
  {\bibinfo  {journal} {Phys. Rept.}\ }\textbf {\bibinfo {volume} {420}},\
  \bibinfo {pages} {1} (\bibinfo {year} {2005})},\ \Eprint
  {https://arxiv.org/abs/hep-ph/0406039} {arXiv:hep-ph/0406039} \BibitemShut
  {NoStop}%
\bibitem [{\citenamefont {Altmannshofer}\ \emph {et~al.}(2017)\citenamefont
  {Altmannshofer}, \citenamefont {Dev},\ and\ \citenamefont
  {Soni}}]{Altmannshofer:2017poe}%
  \BibitemOpen
  \bibfield  {author} {\bibinfo {author} {\bibfnamefont {W.}~\bibnamefont
  {Altmannshofer}}, \bibinfo {author} {\bibfnamefont {P.~S.~B.}\ \bibnamefont
  {Dev}},\ and\ \bibinfo {author} {\bibfnamefont {A.}~\bibnamefont {Soni}},\
  }\bibfield  {title} {\bibinfo {title} {{$R_{D^{(*)}}$ anomaly: A possible
  hint for natural supersymmetry with $R$-parity violation}},\ }\href
  {https://doi.org/10.1103/PhysRevD.96.095010} {\bibfield  {journal} {\bibinfo
  {journal} {Phys. Rev. D}\ }\textbf {\bibinfo {volume} {96}},\ \bibinfo
  {pages} {095010} (\bibinfo {year} {2017})},\ \Eprint
  {https://arxiv.org/abs/1704.06659} {arXiv:1704.06659 [hep-ph]} \BibitemShut
  {NoStop}%
\bibitem [{\citenamefont {Altmannshofer}\ \emph {et~al.}(2020)\citenamefont
  {Altmannshofer}, \citenamefont {Dev}, \citenamefont {Soni},\ and\
  \citenamefont {Sui}}]{Altmannshofer:2020axr}%
  \BibitemOpen
  \bibfield  {author} {\bibinfo {author} {\bibfnamefont {W.}~\bibnamefont
  {Altmannshofer}}, \bibinfo {author} {\bibfnamefont {P.~S.~B.}\ \bibnamefont
  {Dev}}, \bibinfo {author} {\bibfnamefont {A.}~\bibnamefont {Soni}},\ and\
  \bibinfo {author} {\bibfnamefont {Y.}~\bibnamefont {Sui}},\ }\bibfield
  {title} {\bibinfo {title} {{Addressing R$_{D^{(*)}}$, R$_{K^{(*)}}$, muon
  $g-2$ and ANITA anomalies in a minimal $R$-parity violating supersymmetric
  framework}},\ }\href {https://doi.org/10.1103/PhysRevD.102.015031} {\bibfield
   {journal} {\bibinfo  {journal} {Phys. Rev. D}\ }\textbf {\bibinfo {volume}
  {102}},\ \bibinfo {pages} {015031} (\bibinfo {year} {2020})},\ \Eprint
  {https://arxiv.org/abs/2002.12910} {arXiv:2002.12910 [hep-ph]} \BibitemShut
  {NoStop}%
\bibitem [{\citenamefont {Dev}\ \emph {et~al.}(2022)\citenamefont {Dev},
  \citenamefont {Soni},\ and\ \citenamefont {Xu}}]{Dev:2021ipu}%
  \BibitemOpen
  \bibfield  {author} {\bibinfo {author} {\bibfnamefont {P.~S.~B.}\
  \bibnamefont {Dev}}, \bibinfo {author} {\bibfnamefont {A.}~\bibnamefont
  {Soni}},\ and\ \bibinfo {author} {\bibfnamefont {F.}~\bibnamefont {Xu}},\
  }\bibfield  {title} {\bibinfo {title} {{Hints of natural supersymmetry in
  flavor anomalies?}},\ }\href {https://doi.org/10.1103/PhysRevD.106.015014}
  {\bibfield  {journal} {\bibinfo  {journal} {Phys. Rev. D}\ }\textbf {\bibinfo
  {volume} {106}},\ \bibinfo {pages} {015014} (\bibinfo {year} {2022})},\
  \Eprint {https://arxiv.org/abs/2106.15647} {arXiv:2106.15647 [hep-ph]}
  \BibitemShut {NoStop}%
\bibitem [{\citenamefont {Fischer}\ \emph {et~al.}(2022)\citenamefont {Fischer}
  \emph {et~al.}}]{Fischer:2021sqw}%
  \BibitemOpen
  \bibfield  {author} {\bibinfo {author} {\bibfnamefont {O.}~\bibnamefont
  {Fischer}} \emph {et~al.},\ }\bibfield  {title} {\bibinfo {title} {{Unveiling
  hidden physics at the LHC}},\ }\href
  {https://doi.org/10.1140/epjc/s10052-022-10541-4} {\bibfield  {journal}
  {\bibinfo  {journal} {Eur. Phys. J. C}\ }\textbf {\bibinfo {volume} {82}},\
  \bibinfo {pages} {665} (\bibinfo {year} {2022})},\ \Eprint
  {https://arxiv.org/abs/2109.06065} {arXiv:2109.06065 [hep-ph]} \BibitemShut
  {NoStop}%
\bibitem [{\citenamefont {Crivellin}\ and\ \citenamefont
  {Hoferichter}(2021)}]{Crivellin:2021sff}%
  \BibitemOpen
  \bibfield  {author} {\bibinfo {author} {\bibfnamefont {A.}~\bibnamefont
  {Crivellin}}\ and\ \bibinfo {author} {\bibfnamefont {M.}~\bibnamefont
  {Hoferichter}},\ }\bibfield  {title} {\bibinfo {title} {{Hints of lepton
  flavor universality violations}},\ }\href
  {https://doi.org/10.1126/science.abk2450} {\bibfield  {journal} {\bibinfo
  {journal} {Science}\ }\textbf {\bibinfo {volume} {374}},\ \bibinfo {pages}
  {1051} (\bibinfo {year} {2021})},\ \Eprint {https://arxiv.org/abs/2111.12739}
  {arXiv:2111.12739 [hep-ph]} \BibitemShut {NoStop}%
\bibitem [{\citenamefont {Deshpande}\ and\ \citenamefont
  {Menon}(2013)}]{Deshpande:2012rr}%
  \BibitemOpen
  \bibfield  {author} {\bibinfo {author} {\bibfnamefont {N.~G.}\ \bibnamefont
  {Deshpande}}\ and\ \bibinfo {author} {\bibfnamefont {A.}~\bibnamefont
  {Menon}},\ }\bibfield  {title} {\bibinfo {title} {{Hints of R-parity
  violation in B decays into $\tau \nu$}},\ }\href
  {https://doi.org/10.1007/JHEP01(2013)025} {\bibfield  {journal} {\bibinfo
  {journal} {JHEP}\ }\textbf {\bibinfo {volume} {01}},\ \bibinfo {pages}
  {025}},\ \Eprint {https://arxiv.org/abs/1208.4134} {arXiv:1208.4134 [hep-ph]}
  \BibitemShut {NoStop}%
\bibitem [{\citenamefont {Biswas}\ \emph {et~al.}(2015)\citenamefont {Biswas},
  \citenamefont {Chowdhury}, \citenamefont {Han},\ and\ \citenamefont
  {Lee}}]{Biswas:2014gga}%
  \BibitemOpen
  \bibfield  {author} {\bibinfo {author} {\bibfnamefont {S.}~\bibnamefont
  {Biswas}}, \bibinfo {author} {\bibfnamefont {D.}~\bibnamefont {Chowdhury}},
  \bibinfo {author} {\bibfnamefont {S.}~\bibnamefont {Han}},\ and\ \bibinfo
  {author} {\bibfnamefont {S.~J.}\ \bibnamefont {Lee}},\ }\bibfield  {title}
  {\bibinfo {title} {{Explaining the lepton non-universality at the LHCb and
  CMS within a unified framework}},\ }\href
  {https://doi.org/10.1007/JHEP02(2015)142} {\bibfield  {journal} {\bibinfo
  {journal} {JHEP}\ }\textbf {\bibinfo {volume} {02}},\ \bibinfo {pages}
  {142}},\ \Eprint {https://arxiv.org/abs/1409.0882} {arXiv:1409.0882 [hep-ph]}
  \BibitemShut {NoStop}%
\bibitem [{\citenamefont {Zhu}\ \emph {et~al.}(2016)\citenamefont {Zhu},
  \citenamefont {Gan}, \citenamefont {Wang}, \citenamefont {Fan}, \citenamefont
  {Chang},\ and\ \citenamefont {Xu}}]{Zhu:2016xdg}%
  \BibitemOpen
  \bibfield  {author} {\bibinfo {author} {\bibfnamefont {J.}~\bibnamefont
  {Zhu}}, \bibinfo {author} {\bibfnamefont {H.-M.}\ \bibnamefont {Gan}},
  \bibinfo {author} {\bibfnamefont {R.-M.}\ \bibnamefont {Wang}}, \bibinfo
  {author} {\bibfnamefont {Y.-Y.}\ \bibnamefont {Fan}}, \bibinfo {author}
  {\bibfnamefont {Q.}~\bibnamefont {Chang}},\ and\ \bibinfo {author}
  {\bibfnamefont {Y.-G.}\ \bibnamefont {Xu}},\ }\bibfield  {title} {\bibinfo
  {title} {{Probing the R-parity violating supersymmetric effects in the
  exclusive $b \to c\ell^-\bar{\nu}_\ell$ decays}},\ }\href
  {https://doi.org/10.1103/PhysRevD.93.094023} {\bibfield  {journal} {\bibinfo
  {journal} {Phys. Rev. D}\ }\textbf {\bibinfo {volume} {93}},\ \bibinfo
  {pages} {094023} (\bibinfo {year} {2016})},\ \Eprint
  {https://arxiv.org/abs/1602.06491} {arXiv:1602.06491 [hep-ph]} \BibitemShut
  {NoStop}%
\bibitem [{\citenamefont {Deshpande}\ and\ \citenamefont
  {He}(2017)}]{Deshpande:2016yrv}%
  \BibitemOpen
  \bibfield  {author} {\bibinfo {author} {\bibfnamefont {N.~G.}\ \bibnamefont
  {Deshpande}}\ and\ \bibinfo {author} {\bibfnamefont {X.-G.}\ \bibnamefont
  {He}},\ }\bibfield  {title} {\bibinfo {title} {{Consequences of R-parity
  violating interactions for anomalies in $\bar B\to D^{(*)} \tau \bar \nu$ and
  $b\to s \mu^+\mu^-$}},\ }\href
  {https://doi.org/10.1140/epjc/s10052-017-4707-y} {\bibfield  {journal}
  {\bibinfo  {journal} {Eur. Phys. J. C}\ }\textbf {\bibinfo {volume} {77}},\
  \bibinfo {pages} {134} (\bibinfo {year} {2017})},\ \Eprint
  {https://arxiv.org/abs/1608.04817} {arXiv:1608.04817 [hep-ph]} \BibitemShut
  {NoStop}%
\bibitem [{\citenamefont {Das}\ \emph {et~al.}(2017)\citenamefont {Das},
  \citenamefont {Hati}, \citenamefont {Kumar},\ and\ \citenamefont
  {Mahajan}}]{Das:2017kfo}%
  \BibitemOpen
  \bibfield  {author} {\bibinfo {author} {\bibfnamefont {D.}~\bibnamefont
  {Das}}, \bibinfo {author} {\bibfnamefont {C.}~\bibnamefont {Hati}}, \bibinfo
  {author} {\bibfnamefont {G.}~\bibnamefont {Kumar}},\ and\ \bibinfo {author}
  {\bibfnamefont {N.}~\bibnamefont {Mahajan}},\ }\bibfield  {title} {\bibinfo
  {title} {{Scrutinizing $R$-parity violating interactions in light of
  $R_{K^{(\ast)}}$ data}},\ }\href {https://doi.org/10.1103/PhysRevD.96.095033}
  {\bibfield  {journal} {\bibinfo  {journal} {Phys. Rev. D}\ }\textbf {\bibinfo
  {volume} {96}},\ \bibinfo {pages} {095033} (\bibinfo {year} {2017})},\
  \Eprint {https://arxiv.org/abs/1705.09188} {arXiv:1705.09188 [hep-ph]}
  \BibitemShut {NoStop}%
\bibitem [{\citenamefont {Earl}\ and\ \citenamefont
  {Gr\'egoire}(2018)}]{Earl:2018snx}%
  \BibitemOpen
  \bibfield  {author} {\bibinfo {author} {\bibfnamefont {K.}~\bibnamefont
  {Earl}}\ and\ \bibinfo {author} {\bibfnamefont {T.}~\bibnamefont
  {Gr\'egoire}},\ }\bibfield  {title} {\bibinfo {title} {{Contributions to ${b
  \rightarrow s \ell \ell}$ Anomalies from ${R}$-Parity Violating
  Interactions}},\ }\href {https://doi.org/10.1007/JHEP08(2018)201} {\bibfield
  {journal} {\bibinfo  {journal} {JHEP}\ }\textbf {\bibinfo {volume} {08}},\
  \bibinfo {pages} {201}},\ \Eprint {https://arxiv.org/abs/1806.01343}
  {arXiv:1806.01343 [hep-ph]} \BibitemShut {NoStop}%
\bibitem [{\citenamefont {Trifinopoulos}(2018)}]{Trifinopoulos:2018rna}%
  \BibitemOpen
  \bibfield  {author} {\bibinfo {author} {\bibfnamefont {S.}~\bibnamefont
  {Trifinopoulos}},\ }\bibfield  {title} {\bibinfo {title} {{Revisiting
  R-parity violating interactions as an explanation of the B-physics
  anomalies}},\ }\href {https://doi.org/10.1140/epjc/s10052-018-6280-4}
  {\bibfield  {journal} {\bibinfo  {journal} {Eur. Phys. J. C}\ }\textbf
  {\bibinfo {volume} {78}},\ \bibinfo {pages} {803} (\bibinfo {year} {2018})},\
  \Eprint {https://arxiv.org/abs/1807.01638} {arXiv:1807.01638 [hep-ph]}
  \BibitemShut {NoStop}%
\bibitem [{\citenamefont {Hu}\ \emph {et~al.}(2019)\citenamefont {Hu},
  \citenamefont {Li}, \citenamefont {Muramatsu},\ and\ \citenamefont
  {Yang}}]{Hu:2018lmk}%
  \BibitemOpen
  \bibfield  {author} {\bibinfo {author} {\bibfnamefont {Q.-Y.}\ \bibnamefont
  {Hu}}, \bibinfo {author} {\bibfnamefont {X.-Q.}\ \bibnamefont {Li}}, \bibinfo
  {author} {\bibfnamefont {Y.}~\bibnamefont {Muramatsu}},\ and\ \bibinfo
  {author} {\bibfnamefont {Y.-D.}\ \bibnamefont {Yang}},\ }\bibfield  {title}
  {\bibinfo {title} {{R-parity violating solutions to the $R_{D^{(\ast)}}$
  anomaly and their GUT-scale unifications}},\ }\href
  {https://doi.org/10.1103/PhysRevD.99.015008} {\bibfield  {journal} {\bibinfo
  {journal} {Phys. Rev. D}\ }\textbf {\bibinfo {volume} {99}},\ \bibinfo
  {pages} {015008} (\bibinfo {year} {2019})},\ \Eprint
  {https://arxiv.org/abs/1808.01419} {arXiv:1808.01419 [hep-ph]} \BibitemShut
  {NoStop}%
\bibitem [{\citenamefont {Trifinopoulos}(2019)}]{Trifinopoulos:2019lyo}%
  \BibitemOpen
  \bibfield  {author} {\bibinfo {author} {\bibfnamefont {S.}~\bibnamefont
  {Trifinopoulos}},\ }\bibfield  {title} {\bibinfo {title} {{B -physics
  anomalies: The bridge between R -parity violating supersymmetry and flavored
  dark matter}},\ }\href {https://doi.org/10.1103/PhysRevD.100.115022}
  {\bibfield  {journal} {\bibinfo  {journal} {Phys. Rev. D}\ }\textbf {\bibinfo
  {volume} {100}},\ \bibinfo {pages} {115022} (\bibinfo {year} {2019})},\
  \Eprint {https://arxiv.org/abs/1904.12940} {arXiv:1904.12940 [hep-ph]}
  \BibitemShut {NoStop}%
\bibitem [{\citenamefont {Wang}\ \emph {et~al.}(2019)\citenamefont {Wang},
  \citenamefont {Yang},\ and\ \citenamefont {Yuan}}]{Wang:2019trs}%
  \BibitemOpen
  \bibfield  {author} {\bibinfo {author} {\bibfnamefont {D.-Y.}\ \bibnamefont
  {Wang}}, \bibinfo {author} {\bibfnamefont {Y.-D.}\ \bibnamefont {Yang}},\
  and\ \bibinfo {author} {\bibfnamefont {X.-B.}\ \bibnamefont {Yuan}},\
  }\bibfield  {title} {\bibinfo {title} {{$b \to c\tau\bar\nu$ decays in
  supersymmetry with $R$-parity violation}},\ }\href
  {https://doi.org/10.1088/1674-1137/43/8/083103} {\bibfield  {journal}
  {\bibinfo  {journal} {Chin. Phys. C}\ }\textbf {\bibinfo {volume} {43}},\
  \bibinfo {pages} {083103} (\bibinfo {year} {2019})},\ \Eprint
  {https://arxiv.org/abs/1905.08784} {arXiv:1905.08784 [hep-ph]} \BibitemShut
  {NoStop}%
\bibitem [{\citenamefont {Hu}\ and\ \citenamefont {Huang}(2020)}]{Hu:2019ahp}%
  \BibitemOpen
  \bibfield  {author} {\bibinfo {author} {\bibfnamefont {Q.-Y.}\ \bibnamefont
  {Hu}}\ and\ \bibinfo {author} {\bibfnamefont {L.-L.}\ \bibnamefont {Huang}},\
  }\bibfield  {title} {\bibinfo {title} {{Explaining $b\to s \ell^+ \ell^-$
  data by sneutrinos in the $R$ -parity violating MSSM}},\ }\href
  {https://doi.org/10.1103/PhysRevD.101.035030} {\bibfield  {journal} {\bibinfo
   {journal} {Phys. Rev. D}\ }\textbf {\bibinfo {volume} {101}},\ \bibinfo
  {pages} {035030} (\bibinfo {year} {2020})},\ \Eprint
  {https://arxiv.org/abs/1912.03676} {arXiv:1912.03676 [hep-ph]} \BibitemShut
  {NoStop}%
\bibitem [{\citenamefont {Zheng}\ and\ \citenamefont
  {Zhang}(2021)}]{Zheng:2021wnu}%
  \BibitemOpen
  \bibfield  {author} {\bibinfo {author} {\bibfnamefont {M.-D.}\ \bibnamefont
  {Zheng}}\ and\ \bibinfo {author} {\bibfnamefont {H.-H.}\ \bibnamefont
  {Zhang}},\ }\bibfield  {title} {\bibinfo {title} {{Studying the $b\rightarrow
  s \ell^+\ell^-$ anomalies and $(g-2)_{\mu}$ in $R$-parity violating MSSM
  framework with the inverse seesaw mechanism}},\ }\href
  {https://doi.org/10.1103/PhysRevD.104.115023} {\bibfield  {journal} {\bibinfo
   {journal} {Phys. Rev. D}\ }\textbf {\bibinfo {volume} {104}},\ \bibinfo
  {pages} {115023} (\bibinfo {year} {2021})},\ \Eprint
  {https://arxiv.org/abs/2105.06954} {arXiv:2105.06954 [hep-ph]} \BibitemShut
  {NoStop}%
\bibitem [{\citenamefont {Bardhan}\ \emph {et~al.}(2021)\citenamefont
  {Bardhan}, \citenamefont {Ghosh},\ and\ \citenamefont
  {Sachdeva}}]{Bardhan:2021adp}%
  \BibitemOpen
  \bibfield  {author} {\bibinfo {author} {\bibfnamefont {D.}~\bibnamefont
  {Bardhan}}, \bibinfo {author} {\bibfnamefont {D.}~\bibnamefont {Ghosh}},\
  and\ \bibinfo {author} {\bibfnamefont {D.}~\bibnamefont {Sachdeva}},\
  }\bibfield  {title} {\bibinfo {title} {{$R_{K^{(\ast)}}$ from RPV-SUSY
  sneutrinos}},\ }\href@noop {} {\  (\bibinfo {year} {2021})},\ \Eprint
  {https://arxiv.org/abs/2107.10163} {arXiv:2107.10163 [hep-ph]} \BibitemShut
  {NoStop}%
\bibitem [{\citenamefont {Zheng}\ \emph {et~al.}(2022)\citenamefont {Zheng},
  \citenamefont {Chen},\ and\ \citenamefont {Zhang}}]{Zheng:2022ssr}%
  \BibitemOpen
  \bibfield  {author} {\bibinfo {author} {\bibfnamefont {M.-D.}\ \bibnamefont
  {Zheng}}, \bibinfo {author} {\bibfnamefont {F.-Z.}\ \bibnamefont {Chen}},\
  and\ \bibinfo {author} {\bibfnamefont {H.-H.}\ \bibnamefont {Zhang}},\
  }\bibfield  {title} {\bibinfo {title} {{Explaining anomalies of B-physics,
  muon $g-2$ and W mass in R-parity violating MSSM with seesaw mechanism}},\
  }\href {https://doi.org/10.1140/epjc/s10052-022-10822-y} {\bibfield
  {journal} {\bibinfo  {journal} {Eur. Phys. J. C}\ }\textbf {\bibinfo {volume}
  {82}},\ \bibinfo {pages} {895} (\bibinfo {year} {2022})},\ \Eprint
  {https://arxiv.org/abs/2207.07636} {arXiv:2207.07636 [hep-ph]} \BibitemShut
  {NoStop}%
\bibitem [{\citenamefont {Chakraborty}\ and\ \citenamefont
  {Chakraborty}(2016)}]{Chakraborty:2015bsk}%
  \BibitemOpen
  \bibfield  {author} {\bibinfo {author} {\bibfnamefont {A.}~\bibnamefont
  {Chakraborty}}\ and\ \bibinfo {author} {\bibfnamefont {S.}~\bibnamefont
  {Chakraborty}},\ }\bibfield  {title} {\bibinfo {title} {{Probing
  $(g-2)_{\mu}$ at the LHC in the paradigm of $R$-parity violating MSSM}},\
  }\href {https://doi.org/10.1103/PhysRevD.93.075035} {\bibfield  {journal}
  {\bibinfo  {journal} {Phys. Rev. D}\ }\textbf {\bibinfo {volume} {93}},\
  \bibinfo {pages} {075035} (\bibinfo {year} {2016})},\ \Eprint
  {https://arxiv.org/abs/1511.08874} {arXiv:1511.08874 [hep-ph]} \BibitemShut
  {NoStop}%
\bibitem [{ATL(2021)}]{ATLAS:2021eyc}%
  \BibitemOpen
  \bibfield  {title} {\bibinfo {title} {{Search for new phenomena in three- or
  four-lepton events in $pp$ collisions at $\sqrt{s} = $ 13 TeV with the ATLAS
  detector}},\ }\href@noop {} {\  (\bibinfo {year} {2021})},\ \bibinfo {note}
  {aTLAS-CONF-2021-011}\BibitemShut {NoStop}%
\bibitem [{\citenamefont {Decamp}\ \emph {et~al.}(1992)\citenamefont {Decamp}
  \emph {et~al.}}]{ALEPH:1991qhf}%
  \BibitemOpen
  \bibfield  {author} {\bibinfo {author} {\bibfnamefont {D.}~\bibnamefont
  {Decamp}} \emph {et~al.} (\bibinfo {collaboration} {ALEPH}),\ }\bibfield
  {title} {\bibinfo {title} {{Searches for new particles in $Z$ decays using
  the ALEPH detector}},\ }\href {https://doi.org/10.1016/0370-1573(92)90177-2}
  {\bibfield  {journal} {\bibinfo  {journal} {Phys. Rept.}\ }\textbf {\bibinfo
  {volume} {216}},\ \bibinfo {pages} {253} (\bibinfo {year}
  {1992})}\BibitemShut {NoStop}%
\bibitem [{\citenamefont {Aad}\ \emph {et~al.}(2021{\natexlab{a}})\citenamefont
  {Aad} \emph {et~al.}}]{ATLAS:2021yyr}%
  \BibitemOpen
  \bibfield  {author} {\bibinfo {author} {\bibfnamefont {G.}~\bibnamefont
  {Aad}} \emph {et~al.} (\bibinfo {collaboration} {ATLAS}),\ }\bibfield
  {title} {\bibinfo {title} {{Search for supersymmetry in events with four or
  more charged leptons in 139 fb$^{-1}$ of $\sqrt{s}$ = 13 TeV pp collisions
  with the ATLAS detector}},\ }\href {https://doi.org/10.1007/JHEP07(2021)167}
  {\bibfield  {journal} {\bibinfo  {journal} {JHEP}\ }\textbf {\bibinfo
  {volume} {07}},\ \bibinfo {pages} {167}},\ \Eprint
  {https://arxiv.org/abs/2103.11684} {arXiv:2103.11684 [hep-ex]} \BibitemShut
  {NoStop}%
\bibitem [{\citenamefont {Kim}\ \emph {et~al.}(2001)\citenamefont {Kim},
  \citenamefont {Kyae},\ and\ \citenamefont {Lee}}]{Kim:2001se}%
  \BibitemOpen
  \bibfield  {author} {\bibinfo {author} {\bibfnamefont {J.~E.}\ \bibnamefont
  {Kim}}, \bibinfo {author} {\bibfnamefont {B.}~\bibnamefont {Kyae}},\ and\
  \bibinfo {author} {\bibfnamefont {H.~M.}\ \bibnamefont {Lee}},\ }\bibfield
  {title} {\bibinfo {title} {{Effective supersymmetric theory and (g-2)(muon
  with R-parity violation}},\ }\href
  {https://doi.org/10.1016/S0370-2693(01)01134-0} {\bibfield  {journal}
  {\bibinfo  {journal} {Phys. Lett. B}\ }\textbf {\bibinfo {volume} {520}},\
  \bibinfo {pages} {298} (\bibinfo {year} {2001})},\ \Eprint
  {https://arxiv.org/abs/hep-ph/0103054} {arXiv:hep-ph/0103054} \BibitemShut
  {NoStop}%
\bibitem [{\citenamefont {Leveille}(1978)}]{Leveille:1977rc}%
  \BibitemOpen
  \bibfield  {author} {\bibinfo {author} {\bibfnamefont {J.~P.}\ \bibnamefont
  {Leveille}},\ }\bibfield  {title} {\bibinfo {title} {{The Second Order Weak
  Correction to (G-2) of the Muon in Arbitrary Gauge Models}},\ }\href
  {https://doi.org/10.1016/0550-3213(78)90051-2} {\bibfield  {journal}
  {\bibinfo  {journal} {Nucl. Phys. B}\ }\textbf {\bibinfo {volume} {137}},\
  \bibinfo {pages} {63} (\bibinfo {year} {1978})}\BibitemShut {NoStop}%
\bibitem [{\citenamefont {Moroi}(1996)}]{Moroi:1995yh}%
  \BibitemOpen
  \bibfield  {author} {\bibinfo {author} {\bibfnamefont {T.}~\bibnamefont
  {Moroi}},\ }\bibfield  {title} {\bibinfo {title} {{The Muon anomalous
  magnetic dipole moment in the minimal supersymmetric standard model}},\
  }\href {https://doi.org/10.1103/PhysRevD.53.6565} {\bibfield  {journal}
  {\bibinfo  {journal} {Phys. Rev. D}\ }\textbf {\bibinfo {volume} {53}},\
  \bibinfo {pages} {6565} (\bibinfo {year} {1996})},\ \bibinfo {note}
  {[Erratum: Phys.Rev.D 56, 4424 (1997)]},\ \Eprint
  {https://arxiv.org/abs/hep-ph/9512396} {arXiv:hep-ph/9512396} \BibitemShut
  {NoStop}%
\bibitem [{\citenamefont {Baum}\ \emph {et~al.}(2022)\citenamefont {Baum},
  \citenamefont {Carena}, \citenamefont {Shah},\ and\ \citenamefont
  {Wagner}}]{Baum:2021qzx}%
  \BibitemOpen
  \bibfield  {author} {\bibinfo {author} {\bibfnamefont {S.}~\bibnamefont
  {Baum}}, \bibinfo {author} {\bibfnamefont {M.}~\bibnamefont {Carena}},
  \bibinfo {author} {\bibfnamefont {N.~R.}\ \bibnamefont {Shah}},\ and\
  \bibinfo {author} {\bibfnamefont {C.~E.~M.}\ \bibnamefont {Wagner}},\
  }\bibfield  {title} {\bibinfo {title} {{The tiny (g-2) muon wobble from
  small-$\mu$ supersymmetry}},\ }\href
  {https://doi.org/10.1007/JHEP01(2022)025} {\bibfield  {journal} {\bibinfo
  {journal} {JHEP}\ }\textbf {\bibinfo {volume} {01}},\ \bibinfo {pages}
  {025}},\ \Eprint {https://arxiv.org/abs/2104.03302} {arXiv:2104.03302
  [hep-ph]} \BibitemShut {NoStop}%
\bibitem [{\citenamefont {Chakraborti}\ \emph {et~al.}(2022)\citenamefont
  {Chakraborti}, \citenamefont {Iwamoto}, \citenamefont {Kim}, \citenamefont
  {Mase\l{}ek},\ and\ \citenamefont {Sakurai}}]{Chakraborti:2022vds}%
  \BibitemOpen
  \bibfield  {author} {\bibinfo {author} {\bibfnamefont {M.}~\bibnamefont
  {Chakraborti}}, \bibinfo {author} {\bibfnamefont {S.}~\bibnamefont
  {Iwamoto}}, \bibinfo {author} {\bibfnamefont {J.~S.}\ \bibnamefont {Kim}},
  \bibinfo {author} {\bibfnamefont {R.}~\bibnamefont {Mase\l{}ek}},\ and\
  \bibinfo {author} {\bibfnamefont {K.}~\bibnamefont {Sakurai}},\ }\bibfield
  {title} {\bibinfo {title} {{Supersymmetric explanation of the muon g
  \textendash{} 2 anomaly with and without stable neutralino}},\ }\href
  {https://doi.org/10.1007/JHEP08(2022)124} {\bibfield  {journal} {\bibinfo
  {journal} {JHEP}\ }\textbf {\bibinfo {volume} {08}},\ \bibinfo {pages}
  {124}},\ \Eprint {https://arxiv.org/abs/2202.12928} {arXiv:2202.12928
  [hep-ph]} \BibitemShut {NoStop}%
\bibitem [{\citenamefont {Aaboud}\ \emph {et~al.}(2018)\citenamefont {Aaboud}
  \emph {et~al.}}]{ATLAS:2018mrn}%
  \BibitemOpen
  \bibfield  {author} {\bibinfo {author} {\bibfnamefont {M.}~\bibnamefont
  {Aaboud}} \emph {et~al.} (\bibinfo {collaboration} {ATLAS}),\ }\bibfield
  {title} {\bibinfo {title} {{Search for lepton-flavor violation in
  different-flavor, high-mass final states in $pp$ collisions at $\sqrt s=13 $
  TeV with the ATLAS detector}},\ }\href
  {https://doi.org/10.1103/PhysRevD.98.092008} {\bibfield  {journal} {\bibinfo
  {journal} {Phys. Rev. D}\ }\textbf {\bibinfo {volume} {98}},\ \bibinfo
  {pages} {092008} (\bibinfo {year} {2018})},\ \Eprint
  {https://arxiv.org/abs/1807.06573} {arXiv:1807.06573 [hep-ex]} \BibitemShut
  {NoStop}%
\bibitem [{\citenamefont {Fetscher}\ \emph {et~al.}(1986)\citenamefont
  {Fetscher}, \citenamefont {Gerber},\ and\ \citenamefont
  {Johnson}}]{Fetscher:1986uj}%
  \BibitemOpen
  \bibfield  {author} {\bibinfo {author} {\bibfnamefont {W.}~\bibnamefont
  {Fetscher}}, \bibinfo {author} {\bibfnamefont {H.~J.}\ \bibnamefont
  {Gerber}},\ and\ \bibinfo {author} {\bibfnamefont {K.~F.}\ \bibnamefont
  {Johnson}},\ }\bibfield  {title} {\bibinfo {title} {{Muon Decay: Complete
  Determination of the Interaction and Comparison with the Standard Model}},\
  }\href {https://doi.org/10.1016/0370-2693(86)91239-6} {\bibfield  {journal}
  {\bibinfo  {journal} {Phys. Lett. B}\ }\textbf {\bibinfo {volume} {173}},\
  \bibinfo {pages} {102} (\bibinfo {year} {1986})}\BibitemShut {NoStop}%
\bibitem [{\citenamefont {Kuno}\ and\ \citenamefont
  {Okada}(2001)}]{Kuno:1999jp}%
  \BibitemOpen
  \bibfield  {author} {\bibinfo {author} {\bibfnamefont {Y.}~\bibnamefont
  {Kuno}}\ and\ \bibinfo {author} {\bibfnamefont {Y.}~\bibnamefont {Okada}},\
  }\bibfield  {title} {\bibinfo {title} {{Muon decay and physics beyond the
  standard model}},\ }\href {https://doi.org/10.1103/RevModPhys.73.151}
  {\bibfield  {journal} {\bibinfo  {journal} {Rev. Mod. Phys.}\ }\textbf
  {\bibinfo {volume} {73}},\ \bibinfo {pages} {151} (\bibinfo {year} {2001})},\
  \Eprint {https://arxiv.org/abs/hep-ph/9909265} {arXiv:hep-ph/9909265}
  \BibitemShut {NoStop}%
\bibitem [{\citenamefont {Barger}\ \emph {et~al.}(1989)\citenamefont {Barger},
  \citenamefont {Giudice},\ and\ \citenamefont {Han}}]{Barger:1989rk}%
  \BibitemOpen
  \bibfield  {author} {\bibinfo {author} {\bibfnamefont {V.~D.}\ \bibnamefont
  {Barger}}, \bibinfo {author} {\bibfnamefont {G.~F.}\ \bibnamefont
  {Giudice}},\ and\ \bibinfo {author} {\bibfnamefont {T.}~\bibnamefont {Han}},\
  }\bibfield  {title} {\bibinfo {title} {{Some New Aspects of Supersymmetry
  R-Parity Violating Interactions}},\ }\href
  {https://doi.org/10.1103/PhysRevD.40.2987} {\bibfield  {journal} {\bibinfo
  {journal} {Phys. Rev. D}\ }\textbf {\bibinfo {volume} {40}},\ \bibinfo
  {pages} {2987} (\bibinfo {year} {1989})}\BibitemShut {NoStop}%
\bibitem [{\citenamefont {Hall}\ and\ \citenamefont
  {Suzuki}(1984)}]{Hall:1983id}%
  \BibitemOpen
  \bibfield  {author} {\bibinfo {author} {\bibfnamefont {L.~J.}\ \bibnamefont
  {Hall}}\ and\ \bibinfo {author} {\bibfnamefont {M.}~\bibnamefont {Suzuki}},\
  }\bibfield  {title} {\bibinfo {title} {{Explicit R-Parity Breaking in
  Supersymmetric Models}},\ }\href
  {https://doi.org/10.1016/0550-3213(84)90513-3} {\bibfield  {journal}
  {\bibinfo  {journal} {Nucl. Phys. B}\ }\textbf {\bibinfo {volume} {231}},\
  \bibinfo {pages} {419} (\bibinfo {year} {1984})}\BibitemShut {NoStop}%
\bibitem [{\citenamefont {Babu}\ and\ \citenamefont
  {Mohapatra}(1990)}]{Babu:1989px}%
  \BibitemOpen
  \bibfield  {author} {\bibinfo {author} {\bibfnamefont {K.~S.}\ \bibnamefont
  {Babu}}\ and\ \bibinfo {author} {\bibfnamefont {R.~N.}\ \bibnamefont
  {Mohapatra}},\ }\bibfield  {title} {\bibinfo {title} {{Supersymmetry and
  Large Transition Magnetic Moment of the Neutrino}},\ }\href
  {https://doi.org/10.1103/PhysRevLett.64.1705} {\bibfield  {journal} {\bibinfo
   {journal} {Phys. Rev. Lett.}\ }\textbf {\bibinfo {volume} {64}},\ \bibinfo
  {pages} {1705} (\bibinfo {year} {1990})}\BibitemShut {NoStop}%
\bibitem [{\citenamefont {Davidson}\ and\ \citenamefont
  {Losada}(2002)}]{Davidson:2000ne}%
  \BibitemOpen
  \bibfield  {author} {\bibinfo {author} {\bibfnamefont {S.}~\bibnamefont
  {Davidson}}\ and\ \bibinfo {author} {\bibfnamefont {M.}~\bibnamefont
  {Losada}},\ }\bibfield  {title} {\bibinfo {title} {{Basis independent
  neutrino masses in the R(p) violating MSSM}},\ }\href
  {https://doi.org/10.1103/PhysRevD.65.075025} {\bibfield  {journal} {\bibinfo
  {journal} {Phys. Rev. D}\ }\textbf {\bibinfo {volume} {65}},\ \bibinfo
  {pages} {075025} (\bibinfo {year} {2002})},\ \Eprint
  {https://arxiv.org/abs/hep-ph/0010325} {arXiv:hep-ph/0010325} \BibitemShut
  {NoStop}%
\bibitem [{\citenamefont {Alwall}\ \emph {et~al.}(2014)\citenamefont {Alwall},
  \citenamefont {Frederix}, \citenamefont {Frixione}, \citenamefont {Hirschi},
  \citenamefont {Maltoni}, \citenamefont {Mattelaer}, \citenamefont {Shao},
  \citenamefont {Stelzer}, \citenamefont {Torrielli},\ and\ \citenamefont
  {Zaro}}]{Alwall:2014hca}%
  \BibitemOpen
  \bibfield  {author} {\bibinfo {author} {\bibfnamefont {J.}~\bibnamefont
  {Alwall}}, \bibinfo {author} {\bibfnamefont {R.}~\bibnamefont {Frederix}},
  \bibinfo {author} {\bibfnamefont {S.}~\bibnamefont {Frixione}}, \bibinfo
  {author} {\bibfnamefont {V.}~\bibnamefont {Hirschi}}, \bibinfo {author}
  {\bibfnamefont {F.}~\bibnamefont {Maltoni}}, \bibinfo {author} {\bibfnamefont
  {O.}~\bibnamefont {Mattelaer}}, \bibinfo {author} {\bibfnamefont {H.~S.}\
  \bibnamefont {Shao}}, \bibinfo {author} {\bibfnamefont {T.}~\bibnamefont
  {Stelzer}}, \bibinfo {author} {\bibfnamefont {P.}~\bibnamefont {Torrielli}},\
  and\ \bibinfo {author} {\bibfnamefont {M.}~\bibnamefont {Zaro}},\ }\bibfield
  {title} {\bibinfo {title} {{The automated computation of tree-level and
  next-to-leading order differential cross sections, and their matching to
  parton shower simulations}},\ }\href
  {https://doi.org/10.1007/JHEP07(2014)079} {\bibfield  {journal} {\bibinfo
  {journal} {JHEP}\ }\textbf {\bibinfo {volume} {07}},\ \bibinfo {pages}
  {079}},\ \Eprint {https://arxiv.org/abs/1405.0301} {arXiv:1405.0301 [hep-ph]}
  \BibitemShut {NoStop}%
\bibitem [{\citenamefont {Alloul}\ \emph {et~al.}(2014)\citenamefont {Alloul},
  \citenamefont {Christensen}, \citenamefont {Degrande}, \citenamefont {Duhr},\
  and\ \citenamefont {Fuks}}]{Alloul:2013bka}%
  \BibitemOpen
  \bibfield  {author} {\bibinfo {author} {\bibfnamefont {A.}~\bibnamefont
  {Alloul}}, \bibinfo {author} {\bibfnamefont {N.~D.}\ \bibnamefont
  {Christensen}}, \bibinfo {author} {\bibfnamefont {C.}~\bibnamefont
  {Degrande}}, \bibinfo {author} {\bibfnamefont {C.}~\bibnamefont {Duhr}},\
  and\ \bibinfo {author} {\bibfnamefont {B.}~\bibnamefont {Fuks}},\ }\bibfield
  {title} {\bibinfo {title} {{FeynRules 2.0 - A complete toolbox for tree-level
  phenomenology}},\ }\href {https://doi.org/10.1016/j.cpc.2014.04.012}
  {\bibfield  {journal} {\bibinfo  {journal} {Comput. Phys. Commun.}\ }\textbf
  {\bibinfo {volume} {185}},\ \bibinfo {pages} {2250} (\bibinfo {year}
  {2014})},\ \Eprint {https://arxiv.org/abs/1310.1921} {arXiv:1310.1921
  [hep-ph]} \BibitemShut {NoStop}%
\bibitem [{\citenamefont {Ball}\ \emph {et~al.}(2015)\citenamefont {Ball} \emph
  {et~al.}}]{NNPDF:2014otw}%
  \BibitemOpen
  \bibfield  {author} {\bibinfo {author} {\bibfnamefont {R.~D.}\ \bibnamefont
  {Ball}} \emph {et~al.} (\bibinfo {collaboration} {NNPDF}),\ }\bibfield
  {title} {\bibinfo {title} {{Parton distributions for the LHC Run II}},\
  }\href {https://doi.org/10.1007/JHEP04(2015)040} {\bibfield  {journal}
  {\bibinfo  {journal} {JHEP}\ }\textbf {\bibinfo {volume} {04}},\ \bibinfo
  {pages} {040}},\ \Eprint {https://arxiv.org/abs/1410.8849} {arXiv:1410.8849
  [hep-ph]} \BibitemShut {NoStop}%
\bibitem [{\citenamefont {Catani}\ \emph {et~al.}(1993)\citenamefont {Catani},
  \citenamefont {Dokshitzer}, \citenamefont {Seymour},\ and\ \citenamefont
  {Webber}}]{Catani:1993hr}%
  \BibitemOpen
  \bibfield  {author} {\bibinfo {author} {\bibfnamefont {S.}~\bibnamefont
  {Catani}}, \bibinfo {author} {\bibfnamefont {Y.~L.}\ \bibnamefont
  {Dokshitzer}}, \bibinfo {author} {\bibfnamefont {M.~H.}\ \bibnamefont
  {Seymour}},\ and\ \bibinfo {author} {\bibfnamefont {B.~R.}\ \bibnamefont
  {Webber}},\ }\bibfield  {title} {\bibinfo {title} {{Longitudinally invariant
  $K_t$ clustering algorithms for hadron hadron collisions}},\ }\href
  {https://doi.org/10.1016/0550-3213(93)90166-M} {\bibfield  {journal}
  {\bibinfo  {journal} {Nucl. Phys. B}\ }\textbf {\bibinfo {volume} {406}},\
  \bibinfo {pages} {187} (\bibinfo {year} {1993})}\BibitemShut {NoStop}%
\bibitem [{\citenamefont {Mrenna}\ and\ \citenamefont
  {Skands}(2016)}]{Mrenna:2016sih}%
  \BibitemOpen
  \bibfield  {author} {\bibinfo {author} {\bibfnamefont {S.}~\bibnamefont
  {Mrenna}}\ and\ \bibinfo {author} {\bibfnamefont {P.}~\bibnamefont
  {Skands}},\ }\bibfield  {title} {\bibinfo {title} {{Automated Parton-Shower
  Variations in Pythia 8}},\ }\href
  {https://doi.org/10.1103/PhysRevD.94.074005} {\bibfield  {journal} {\bibinfo
  {journal} {Phys. Rev. D}\ }\textbf {\bibinfo {volume} {94}},\ \bibinfo
  {pages} {074005} (\bibinfo {year} {2016})},\ \Eprint
  {https://arxiv.org/abs/1605.08352} {arXiv:1605.08352 [hep-ph]} \BibitemShut
  {NoStop}%
\bibitem [{\citenamefont {Mangano}\ \emph {et~al.}(2007)\citenamefont
  {Mangano}, \citenamefont {Moretti}, \citenamefont {Piccinini},\ and\
  \citenamefont {Treccani}}]{Mangano:2006rw}%
  \BibitemOpen
  \bibfield  {author} {\bibinfo {author} {\bibfnamefont {M.~L.}\ \bibnamefont
  {Mangano}}, \bibinfo {author} {\bibfnamefont {M.}~\bibnamefont {Moretti}},
  \bibinfo {author} {\bibfnamefont {F.}~\bibnamefont {Piccinini}},\ and\
  \bibinfo {author} {\bibfnamefont {M.}~\bibnamefont {Treccani}},\ }\bibfield
  {title} {\bibinfo {title} {{Matching matrix elements and shower evolution for
  top-quark production in hadronic collisions}},\ }\href
  {https://doi.org/10.1088/1126-6708/2007/01/013} {\bibfield  {journal}
  {\bibinfo  {journal} {JHEP}\ }\textbf {\bibinfo {volume} {01}},\ \bibinfo
  {pages} {013}},\ \Eprint {https://arxiv.org/abs/hep-ph/0611129}
  {arXiv:hep-ph/0611129} \BibitemShut {NoStop}%
\bibitem [{\citenamefont {de~Favereau}\ \emph {et~al.}(2014)\citenamefont
  {de~Favereau}, \citenamefont {Delaere}, \citenamefont {Demin}, \citenamefont
  {Giammanco}, \citenamefont {Lema\^\i{}tre}, \citenamefont {Mertens},\ and\
  \citenamefont {Selvaggi}}]{deFavereau:2013fsa}%
  \BibitemOpen
  \bibfield  {author} {\bibinfo {author} {\bibfnamefont {J.}~\bibnamefont
  {de~Favereau}}, \bibinfo {author} {\bibfnamefont {C.}~\bibnamefont
  {Delaere}}, \bibinfo {author} {\bibfnamefont {P.}~\bibnamefont {Demin}},
  \bibinfo {author} {\bibfnamefont {A.}~\bibnamefont {Giammanco}}, \bibinfo
  {author} {\bibfnamefont {V.}~\bibnamefont {Lema\^\i{}tre}}, \bibinfo {author}
  {\bibfnamefont {A.}~\bibnamefont {Mertens}},\ and\ \bibinfo {author}
  {\bibfnamefont {M.}~\bibnamefont {Selvaggi}} (\bibinfo {collaboration}
  {DELPHES 3}),\ }\bibfield  {title} {\bibinfo {title} {{DELPHES 3, A modular
  framework for fast simulation of a generic collider experiment}},\ }\href
  {https://doi.org/10.1007/JHEP02(2014)057} {\bibfield  {journal} {\bibinfo
  {journal} {JHEP}\ }\textbf {\bibinfo {volume} {02}},\ \bibinfo {pages}
  {057}},\ \Eprint {https://arxiv.org/abs/1307.6346} {arXiv:1307.6346 [hep-ex]}
  \BibitemShut {NoStop}%
\bibitem [{\citenamefont {Cacciari}\ \emph {et~al.}(2008)\citenamefont
  {Cacciari}, \citenamefont {Salam},\ and\ \citenamefont
  {Soyez}}]{Cacciari:2008gp}%
  \BibitemOpen
  \bibfield  {author} {\bibinfo {author} {\bibfnamefont {M.}~\bibnamefont
  {Cacciari}}, \bibinfo {author} {\bibfnamefont {G.~P.}\ \bibnamefont
  {Salam}},\ and\ \bibinfo {author} {\bibfnamefont {G.}~\bibnamefont {Soyez}},\
  }\bibfield  {title} {\bibinfo {title} {{The anti-$k_t$ jet clustering
  algorithm}},\ }\href {https://doi.org/10.1088/1126-6708/2008/04/063}
  {\bibfield  {journal} {\bibinfo  {journal} {JHEP}\ }\textbf {\bibinfo
  {volume} {04}},\ \bibinfo {pages} {063}},\ \Eprint
  {https://arxiv.org/abs/0802.1189} {arXiv:0802.1189 [hep-ph]} \BibitemShut
  {NoStop}%
\bibitem [{\citenamefont {Cacciari}\ \emph {et~al.}(2012)\citenamefont
  {Cacciari}, \citenamefont {Salam},\ and\ \citenamefont
  {Soyez}}]{Cacciari:2011ma}%
  \BibitemOpen
  \bibfield  {author} {\bibinfo {author} {\bibfnamefont {M.}~\bibnamefont
  {Cacciari}}, \bibinfo {author} {\bibfnamefont {G.~P.}\ \bibnamefont
  {Salam}},\ and\ \bibinfo {author} {\bibfnamefont {G.}~\bibnamefont {Soyez}},\
  }\bibfield  {title} {\bibinfo {title} {{FastJet User Manual}},\ }\href
  {https://doi.org/10.1140/epjc/s10052-012-1896-2} {\bibfield  {journal}
  {\bibinfo  {journal} {Eur. Phys. J. C}\ }\textbf {\bibinfo {volume} {72}},\
  \bibinfo {pages} {1896} (\bibinfo {year} {2012})},\ \Eprint
  {https://arxiv.org/abs/1111.6097} {arXiv:1111.6097 [hep-ph]} \BibitemShut
  {NoStop}%
\bibitem [{\citenamefont {Cacciari}\ and\ \citenamefont
  {Salam}(2006)}]{Cacciari:2005hq}%
  \BibitemOpen
  \bibfield  {author} {\bibinfo {author} {\bibfnamefont {M.}~\bibnamefont
  {Cacciari}}\ and\ \bibinfo {author} {\bibfnamefont {G.~P.}\ \bibnamefont
  {Salam}},\ }\bibfield  {title} {\bibinfo {title} {{Dispelling the $N^{3}$
  myth for the $k_t$ jet-finder}},\ }\href
  {https://doi.org/10.1016/j.physletb.2006.08.037} {\bibfield  {journal}
  {\bibinfo  {journal} {Phys. Lett. B}\ }\textbf {\bibinfo {volume} {641}},\
  \bibinfo {pages} {57} (\bibinfo {year} {2006})},\ \Eprint
  {https://arxiv.org/abs/hep-ph/0512210} {arXiv:hep-ph/0512210} \BibitemShut
  {NoStop}%
\bibitem [{ATL(2015{\natexlab{a}})}]{ATL-PHYS-PUB-2015-022}%
  \BibitemOpen
  \href {http://cds.cern.ch/record/2037697} {\emph {\bibinfo {title} {{Expected
  performance of the ATLAS $b$-tagging algorithms in Run-2}}}},\ \bibinfo
  {type} {Tech. Rep.}\ \bibinfo {number} {ATL-PHYS-PUB-2015-022}\ (\bibinfo
  {institution} {CERN},\ \bibinfo {address} {Geneva},\ \bibinfo {year}
  {2015})\BibitemShut {NoStop}%
\bibitem [{ATL(2015{\natexlab{b}})}]{ATL-PHYS-PUB-2015-045}%
  \BibitemOpen
  \href {https://cds.cern.ch/record/2064383} {\emph {\bibinfo {title}
  {{Reconstruction, Energy Calibration, and Identification of Hadronically
  Decaying Tau Leptons in the ATLAS Experiment for Run-2 of the LHC}}}},\
  \bibinfo {type} {Tech. Rep.}\ (\bibinfo  {institution} {CERN},\ \bibinfo
  {address} {Geneva},\ \bibinfo {year} {2015})\BibitemShut {NoStop}%
\bibitem [{\citenamefont {Verkerke}\ and\ \citenamefont
  {Kirkby}(2003)}]{Verkerke:2003ir}%
  \BibitemOpen
  \bibfield  {author} {\bibinfo {author} {\bibfnamefont {W.}~\bibnamefont
  {Verkerke}}\ and\ \bibinfo {author} {\bibfnamefont {D.~P.}\ \bibnamefont
  {Kirkby}},\ }\bibfield  {title} {\bibinfo {title} {{The RooFit toolkit for
  data modeling}},\ }\href@noop {} {\bibfield  {journal} {\bibinfo  {journal}
  {eConf}\ }\textbf {\bibinfo {volume} {C0303241}},\ \bibinfo {pages} {MOLT007}
  (\bibinfo {year} {2003})},\ \Eprint {https://arxiv.org/abs/physics/0306116}
  {arXiv:physics/0306116} \BibitemShut {NoStop}%
\bibitem [{\citenamefont {Aad}\ \emph {et~al.}(2014)\citenamefont {Aad} \emph
  {et~al.}}]{ATLAS:2014pjz}%
  \BibitemOpen
  \bibfield  {author} {\bibinfo {author} {\bibfnamefont {G.}~\bibnamefont
  {Aad}} \emph {et~al.} (\bibinfo {collaboration} {ATLAS}),\ }\bibfield
  {title} {\bibinfo {title} {{Search for supersymmetry in events with four or
  more leptons in $\sqrt{s}$ = 8 TeV pp collisions with the ATLAS detector}},\
  }\href {https://doi.org/10.1103/PhysRevD.90.052001} {\bibfield  {journal}
  {\bibinfo  {journal} {Phys. Rev. D}\ }\textbf {\bibinfo {volume} {90}},\
  \bibinfo {pages} {052001} (\bibinfo {year} {2014})},\ \Eprint
  {https://arxiv.org/abs/1405.5086} {arXiv:1405.5086 [hep-ex]} \BibitemShut
  {NoStop}%
\bibitem [{\citenamefont {Aad}\ \emph {et~al.}(2021{\natexlab{b}})\citenamefont
  {Aad} \emph {et~al.}}]{ATLAS:2021kxv}%
  \BibitemOpen
  \bibfield  {author} {\bibinfo {author} {\bibfnamefont {G.}~\bibnamefont
  {Aad}} \emph {et~al.} (\bibinfo {collaboration} {ATLAS}),\ }\bibfield
  {title} {\bibinfo {title} {{Search for new phenomena in events with an
  energetic jet and missing transverse momentum in $pp$ collisions at $\sqrt
  {s}$ =13 TeV with the ATLAS detector}},\ }\href
  {https://doi.org/10.1103/PhysRevD.103.112006} {\bibfield  {journal} {\bibinfo
   {journal} {Phys. Rev. D}\ }\textbf {\bibinfo {volume} {103}},\ \bibinfo
  {pages} {112006} (\bibinfo {year} {2021}{\natexlab{b}})},\ \Eprint
  {https://arxiv.org/abs/2102.10874} {arXiv:2102.10874 [hep-ex]} \BibitemShut
  {NoStop}%
\bibitem [{Ele(2003)}]{Electroweak:2003ram}%
  \BibitemOpen
  \bibfield  {title} {\bibinfo {title} {{A Combination of preliminary
  electroweak measurements and constraints on the standard model}},\
  }\href@noop {} {\  (\bibinfo {year} {2003})},\ \Eprint
  {https://arxiv.org/abs/hep-ex/0312023} {arXiv:hep-ex/0312023} \BibitemShut
  {NoStop}%
\bibitem [{\citenamefont {Achard}\ \emph {et~al.}(2004)\citenamefont {Achard}
  \emph {et~al.}}]{L3:2003yon}%
  \BibitemOpen
  \bibfield  {author} {\bibinfo {author} {\bibfnamefont {P.}~\bibnamefont
  {Achard}} \emph {et~al.} (\bibinfo {collaboration} {L3}),\ }\bibfield
  {title} {\bibinfo {title} {{Single photon and multiphoton events with missing
  energy in $e^{+} e^{-}$ collisions at LEP}},\ }\href
  {https://doi.org/10.1016/j.physletb.2004.01.010} {\bibfield  {journal}
  {\bibinfo  {journal} {Phys. Lett. B}\ }\textbf {\bibinfo {volume} {587}},\
  \bibinfo {pages} {16} (\bibinfo {year} {2004})},\ \Eprint
  {https://arxiv.org/abs/hep-ex/0402002} {arXiv:hep-ex/0402002} \BibitemShut
  {NoStop}%
\bibitem [{\citenamefont {Haber}\ and\ \citenamefont
  {Wyler}(1989)}]{Haber:1988px}%
  \BibitemOpen
  \bibfield  {author} {\bibinfo {author} {\bibfnamefont {H.~E.}\ \bibnamefont
  {Haber}}\ and\ \bibinfo {author} {\bibfnamefont {D.}~\bibnamefont {Wyler}},\
  }\bibfield  {title} {\bibinfo {title} {{RADIATIVE NEUTRALINO DECAY}},\ }\href
  {https://doi.org/10.1016/0550-3213(89)90143-0} {\bibfield  {journal}
  {\bibinfo  {journal} {Nucl. Phys. B}\ }\textbf {\bibinfo {volume} {323}},\
  \bibinfo {pages} {267} (\bibinfo {year} {1989})}\BibitemShut {NoStop}%
\bibitem [{\citenamefont {Dreiner}\ \emph {et~al.}(2023)\citenamefont
  {Dreiner}, \citenamefont {K\"ohler}, \citenamefont {Nangia},\ and\
  \citenamefont {Wang}}]{Dreiner:2022swd}%
  \BibitemOpen
  \bibfield  {author} {\bibinfo {author} {\bibfnamefont {H.~K.}\ \bibnamefont
  {Dreiner}}, \bibinfo {author} {\bibfnamefont {D.}~\bibnamefont {K\"ohler}},
  \bibinfo {author} {\bibfnamefont {S.}~\bibnamefont {Nangia}},\ and\ \bibinfo
  {author} {\bibfnamefont {Z.~S.}\ \bibnamefont {Wang}},\ }\bibfield  {title}
  {\bibinfo {title} {{Searching for a single photon from lightest neutralino
  decays in R-parity-violating supersymmetry at FASER}},\ }\href
  {https://doi.org/10.1007/JHEP02(2023)120} {\bibfield  {journal} {\bibinfo
  {journal} {JHEP}\ }\textbf {\bibinfo {volume} {02}},\ \bibinfo {pages}
  {120}},\ \Eprint {https://arxiv.org/abs/2207.05100} {arXiv:2207.05100
  [hep-ph]} \BibitemShut {NoStop}%
\bibitem [{\citenamefont {Covi}\ \emph {et~al.}(2009)\citenamefont {Covi},
  \citenamefont {Hasenkamp}, \citenamefont {Pokorski},\ and\ \citenamefont
  {Roberts}}]{Covi:2009bk}%
  \BibitemOpen
  \bibfield  {author} {\bibinfo {author} {\bibfnamefont {L.}~\bibnamefont
  {Covi}}, \bibinfo {author} {\bibfnamefont {J.}~\bibnamefont {Hasenkamp}},
  \bibinfo {author} {\bibfnamefont {S.}~\bibnamefont {Pokorski}},\ and\
  \bibinfo {author} {\bibfnamefont {J.}~\bibnamefont {Roberts}},\ }\bibfield
  {title} {\bibinfo {title} {{Gravitino Dark Matter and general neutralino
  NLSP}},\ }\href {https://doi.org/10.1088/1126-6708/2009/11/003} {\bibfield
  {journal} {\bibinfo  {journal} {JHEP}\ }\textbf {\bibinfo {volume} {11}},\
  \bibinfo {pages} {003}},\ \Eprint {https://arxiv.org/abs/0908.3399}
  {arXiv:0908.3399 [hep-ph]} \BibitemShut {NoStop}%
\bibitem [{LHC(2022{\natexlab{a}})}]{LHCb:2022qnv}%
  \BibitemOpen
  \bibfield  {title} {\bibinfo {title} {{Test of lepton universality in $b
  \rightarrow s \ell^+ \ell^-$ decays}},\ }\href@noop {} {\  (\bibinfo {year}
  {2022}{\natexlab{a}})},\ \Eprint {https://arxiv.org/abs/2212.09152}
  {arXiv:2212.09152 [hep-ex]} \BibitemShut {NoStop}%
\bibitem [{LHC(2022{\natexlab{b}})}]{LHCb:2022zom}%
  \BibitemOpen
  \bibfield  {title} {\bibinfo {title} {{Measurement of lepton universality
  parameters in $B^+\to K^+\ell^+\ell^-$ and $B^0\to K^{*0}\ell^+\ell^-$
  decays}},\ }\href@noop {} {\  (\bibinfo {year} {2022}{\natexlab{b}})},\
  \Eprint {https://arxiv.org/abs/2212.09153} {arXiv:2212.09153 [hep-ex]}
  \BibitemShut {NoStop}%
\bibitem [{\citenamefont {Abe}\ \emph {et~al.}(2019)\citenamefont {Abe} \emph
  {et~al.}}]{Abe:2019thb}%
  \BibitemOpen
  \bibfield  {author} {\bibinfo {author} {\bibfnamefont {M.}~\bibnamefont
  {Abe}} \emph {et~al.},\ }\bibfield  {title} {\bibinfo {title} {{A New
  Approach for Measuring the Muon Anomalous Magnetic Moment and Electric Dipole
  Moment}},\ }\href {https://doi.org/10.1093/ptep/ptz030} {\bibfield  {journal}
  {\bibinfo  {journal} {PTEP}\ }\textbf {\bibinfo {volume} {2019}},\ \bibinfo
  {pages} {053C02} (\bibinfo {year} {2019})},\ \Eprint
  {https://arxiv.org/abs/1901.03047} {arXiv:1901.03047 [physics.ins-det]}
  \BibitemShut {NoStop}%
\bibitem [{\citenamefont {Colangelo}\ \emph
  {et~al.}(2022{\natexlab{b}})\citenamefont {Colangelo} \emph
  {et~al.}}]{Colangelo:2022jxc}%
  \BibitemOpen
  \bibfield  {author} {\bibinfo {author} {\bibfnamefont {G.}~\bibnamefont
  {Colangelo}} \emph {et~al.},\ }\bibfield  {title} {\bibinfo {title}
  {{Prospects for precise predictions of $a_\mu$ in the Standard Model}},\
  }\href@noop {} {\  (\bibinfo {year} {2022}{\natexlab{b}})},\ \Eprint
  {https://arxiv.org/abs/2203.15810} {arXiv:2203.15810 [hep-ph]} \BibitemShut
  {NoStop}%
\bibitem [{\citenamefont {Abbiendi}\ \emph {et~al.}(2017)\citenamefont
  {Abbiendi} \emph {et~al.}}]{Abbiendi:2016xup}%
  \BibitemOpen
  \bibfield  {author} {\bibinfo {author} {\bibfnamefont {G.}~\bibnamefont
  {Abbiendi}} \emph {et~al.},\ }\bibfield  {title} {\bibinfo {title}
  {{Measuring the leading hadronic contribution to the muon g-2 via $\mu e$
  scattering}},\ }\href {https://doi.org/10.1140/epjc/s10052-017-4633-z}
  {\bibfield  {journal} {\bibinfo  {journal} {Eur. Phys. J. C}\ }\textbf
  {\bibinfo {volume} {77}},\ \bibinfo {pages} {139} (\bibinfo {year} {2017})},\
  \Eprint {https://arxiv.org/abs/1609.08987} {arXiv:1609.08987 [hep-ex]}
  \BibitemShut {NoStop}%
\bibitem [{\citenamefont {Dev}\ \emph {et~al.}(2020)\citenamefont {Dev},
  \citenamefont {Rodejohann}, \citenamefont {Xu},\ and\ \citenamefont
  {Zhang}}]{Dev:2020drf}%
  \BibitemOpen
  \bibfield  {author} {\bibinfo {author} {\bibfnamefont {P.~S.~B.}\
  \bibnamefont {Dev}}, \bibinfo {author} {\bibfnamefont {W.}~\bibnamefont
  {Rodejohann}}, \bibinfo {author} {\bibfnamefont {X.-J.}\ \bibnamefont {Xu}},\
  and\ \bibinfo {author} {\bibfnamefont {Y.}~\bibnamefont {Zhang}},\ }\bibfield
   {title} {\bibinfo {title} {{MUonE sensitivity to new physics explanations of
  the muon anomalous magnetic moment}},\ }\href
  {https://doi.org/10.1007/JHEP05(2020)053} {\bibfield  {journal} {\bibinfo
  {journal} {JHEP}\ }\textbf {\bibinfo {volume} {05}},\ \bibinfo {pages}
  {053}},\ \Eprint {https://arxiv.org/abs/2002.04822} {arXiv:2002.04822
  [hep-ph]} \BibitemShut {NoStop}%
\bibitem [{\citenamefont {Masiero}\ \emph {et~al.}(2020)\citenamefont
  {Masiero}, \citenamefont {Paradisi},\ and\ \citenamefont
  {Passera}}]{Masiero:2020vxk}%
  \BibitemOpen
  \bibfield  {author} {\bibinfo {author} {\bibfnamefont {A.}~\bibnamefont
  {Masiero}}, \bibinfo {author} {\bibfnamefont {P.}~\bibnamefont {Paradisi}},\
  and\ \bibinfo {author} {\bibfnamefont {M.}~\bibnamefont {Passera}},\
  }\bibfield  {title} {\bibinfo {title} {{New physics at the MUonE experiment
  at CERN}},\ }\href {https://doi.org/10.1103/PhysRevD.102.075013} {\bibfield
  {journal} {\bibinfo  {journal} {Phys. Rev. D}\ }\textbf {\bibinfo {volume}
  {102}},\ \bibinfo {pages} {075013} (\bibinfo {year} {2020})},\ \Eprint
  {https://arxiv.org/abs/2002.05418} {arXiv:2002.05418 [hep-ph]} \BibitemShut
  {NoStop}%
\end{thebibliography}%

\end{document}